\documentclass[aps,prl,superscriptaddress,twocolumn]{revtex4-1}
\usepackage{graphicx}
\usepackage[dvipsnames]{xcolor}
\usepackage{siunitx}
\usepackage{hyperref}
\usepackage{xfrac}
\usepackage{soul}
\usepackage{float}
\usepackage{color}
\usepackage{stmaryrd}
\usepackage{mathtools}
\usepackage[nameinlink,capitalise]{cleveref}
\hypersetup{
colorlinks=true,
linkcolor=blue,
filecolor=magenta,
urlcolor=cyan,
}

\makeatletter
\AtBeginDocument{\let\LS@rot\@undefined}
\makeatother
\begin{document}
\title{Growth and characterization of thorium-doped calcium fluoride single crystals}

\author{Kjeld Beeks}
\author{Tomas Sikorsky}
\author{Veronika Rosecker}
\author{Martin Pressler}
\author{Fabian Schaden}
\author{David Werban}
\author{Niyusha Hosseini}
\author{Lukas Rudischer}
\author{Felix Schneider}

\affiliation{Institute for Atomic and Subatomic Physics, TU Wien, Stadionallee 2, 1020, Vienna, Austria}
\author{Patrick Berwian}
\author{Jochen Friedrich}

\affiliation{Fraunhofer-Institut für Integrierte Systeme
und Bauelementetechnologie IISB, Schottkystraße 10, 91058 Erlangen, Germany}

\author{Dieter Hainz}

\affiliation{TRIGA Center Atominstitut, TU Wien, Stadionallee 2, 1020, Vienna, Austria}

\author{Jan Welch}
\author{Johannes H. Sterba}

\affiliation{CLIP, TRIGA Center Atominstitut TU Wien, Stadionallee 2, 1020, Vienna, Austria}

\author{Georgy Kazakov}
\author{Thorsten Schumm}

\affiliation{Institute for Atomic and Subatomic Physics, TU Wien, Stadionallee 2, 1020, Vienna, Austria}

\date{\today}
\begin{abstract}
We have grown $^{232}$Th:CaF$_2$ and $^{229}$Th:CaF$_2$ single crystals for investigations on the VUV laser-accessible first nuclear excited state of $^{229}$Th. To reach high doping concentrations despite the extreme scarcity (and radioactivity) of $^{229}$Th, we have scaled down the crystal volume by a factor 100 compared to established commercial or scientific growth processes. We use the vertical gradient freeze method on 3.2\,mm diameter seed single crystals with a 2\,mm drilled pocket, filled with a co-precipitated CaF$_2$:ThF$_4$:PbF$_2$ powder in order to grow single crystals. 
Concentrations of $4\cdot10^{19}$\,cm$^{-3}$ have been realized with $^{232}$Th with good ($>$10\%) VUV transmission. However, the intrinsic radioactivity of $^{229}$Th drives radio-induced dissociation during growth and radiation damage after solidification. Both lead to a degradation of VUV transmission, limiting the $^{229}$Th concentration to $<5\cdot10^{17}$\,cm$^{-3}$.


\end{abstract}

\maketitle

\subsection{Introduction}
The radioisotope thorium-229 has a unique nuclear structure in which the first excited state is long-lived and exceptionally low in energy: few electron volts (eV), instead of the common keV-MeV range for nuclear excited states~\cite{von2017direct}. The radiative lifetime of this isomeric state ($^{229m}$Th) is expected to exceed 1000\,seconds~\cite{Tkalya2015} for the bare nucleus. Owing to its low energy in the range of electronic shell transitions, an interaction between the nucleus and its chemical environment is expected~\cite{tkalya2011proposal,Borisyuk2019,Bilous2017, Nickerson2020color,nickerson2021driven}. Studying the interaction of the nucleus with its chemical surrounding presents a unique research opportunity. 

The $^{229}$Th nucleus has attracted many ideas for applications~\cite{tkalya1996processes}, most of them based on nuclear laser spectroscopy. Our main interest is to perform optical nuclear spectroscopy of vacuum ultraviolet (VUV) transparent single crystals containing $^{229}$Th as a dopant~\cite{dessovic2014229thorium}.

The energy of this nuclear isomer state was indirectly measured only recently by two independent methods~\cite{Seiferle2019,sikorsky2020measurement} to $8.15\pm0.45$\,eV (averaged). This corresponds to a wavelength of $152\pm8$\,nm, which is in the VUV range and thus absorbed in air. The measurements of the isomer energy used internal conversion (IC) and gamma emissions from the nucleus respectively. Internal conversion is a common nuclear decay process, where the energy of the excited nucleus is transferred to a shell electron which is ejected if the decay energy exceeds the binding energy. The internal conversion decay channel can have a dramatically different lifetime compared to the radiative decay. The IC lifetime of $7(1)\,\mu$s was measured for neutral $^{229m}$Th on a metal surface~\cite{VonderWense2016}.

To exploit the many prospects of the $^{229}$Th system, internal conversion and other non-radiative decay channels need to be suppressed. For solid-state approaches, this requires the bandgap of the $^{229}$Th-doped crystal material to exceed the isomer excitation energy. Hehlen et al. categorized, which large bandgap materials would be suitable, pointing out the relevance of fluoride crystals~\cite{HEHLEN201391}. Excitation and fluorescence of commercially grown, $^{232}$Th-containing crystals was explored by~\cite{Rellergert_2010} (Th:NaYF, Th:YLF, Th:LiCAF, Na$_2$ThF$_6$, Th:LiSAF) to investigate VUV irradiation induced background and optical transparency.  


The approach in our laboratory is to use CaF$_2$ crystals with an 11.8\,eV direct bandgap~\cite{caf2bandgap1976}. The cutoff of this material is however dominated by a broad indirect exciton bound state at 11.2\,eV~\cite{CaF2opticaltransmission} which diminishes the VUV transmission for photons with energy above $~$9.8\,eV or wavelength below 126\,nm. These exciton states are higher in energy than the 8.15 eV isomer energy and thus non-radiative de-excitation should be prevented. 

Crystal doping will ensure a high amount of addressable nuclei, of the order of 10$^{19}$\,cm$^{-3}$. The presence of the dopant will however modify the band structure of the host crystal and lead to additional electronic defect states~\cite{dessovic2014229thorium}. Interactions between the nucleus and the local crystal fields will lead to line shifts and broadenings~\cite{Kazakov_2012}.


In this work, we describe the in-house growth and characterization of $^{229}$Th-doped CaF$_2$ single crystals. Severe challenges are connected with the inherent radioactivity of the dopant (on the order of~10$^6$\,Bq), its extreme scarcity (milligrams) and general purity requirements of the used materials. To accommodate these, we developed a modified vertical gradient freeze (VGF) machine to grow small-volume crystals ($< 0.1$\,cm$^{-3}$) in vacuum with minimal losses. We demonstrate that the doped material maintains VUV transparency in the range of the expected isomer excitation wavelength and investigate the relation between VUV transmission and the effect of different doping concentrations on both surface and bulk absorption. These crystals are currently used in several attempts to excite the $^{229}$Th using X-ray irradiation \cite{Masuda2019} or VUV irradiation \cite{Jeet2015,Stellmer2018}.


\subsection{Preparation of \textsuperscript{229}ThF\textsubscript{4}:PbF\textsubscript{2}:CaF\textsubscript{2} growth material}
In an exemplary fashion, the following describes the preparation of 45\,mg of $^{229}$ThF$_4$:PbF$_2$:CaF$_2$ powder for growing 3 crystals (15\,mg each) with 3.2(1)\,mm diameter and 11(1)\,mm length. PbF$_2$ acts as a scavenger for oxygen removal (see below) and as a carrier that facilitates the handling of the minuscule (micrograms) amounts of $^{229}$ThF$_4$ during the wet chemistry preparation process. 


$^{229}$Th (7.9\,MBq, Oak Ridge National Laboratory, in dried nitrate form) was dissolved in 0.1 M HNO$_3$ Suprapure grade (Sigma Aldrich) prior to use. All reagents CaF$_2$ (Alfa Aesar), Pb(NO$_3$)$_2$ (Sigma Aldrich), PbF$_2$ (Alfa Aesar), 40\,\% HF (Sigma Aldrich) were purchased from commercial suppliers in trace metal grade and were used as received. Water was purified in-house by triple distillation. Using higher quality CaF$_2$ powder increased the VUV transmission of the grown crystals. 

In a centrifugation vial, lead(II) nitrate (2.9\,mg) was added to and dissolved in a solution of $^{229}$Th in 0.1 M HNO$_3$ (9\,mL, 5.5\,MBq). Subsequently, $^{229}$ThF$_4$:PbF$_2$ was precipitated by addition of hydrofluoric acid (40\,\%, 1\,mL). A white precipitate appeared immediately and was allowed to rest overnight. The supernatant was carefully removed using a pipette after centrifugation and the precipitate was washed with triple distilled water (2\,mL, 6 times). After the fourth washing step, the supernatant was tested for remaining free fluoride ions by adding a small portion of an aqueous solution of CaCl$_2$. No appearance of any white material confirmed the absence of free fluoride ions, and two additional washing steps were performed.

The $^{229}$ThF$_4$:PbF$_2$ was then poured into an aluminum container to avoid powder sticking to the wall, reducing losses when the powder is transferred into the crystal growing machine later. A small portion of water was added and the whole was then dried in an oven at 80\,\textcelsius{} until weight was constant (4\,days). Then CaF$_2$ (28.3\,mg) was added to the $^{229}$ThF$_4$:PbF$_2$, mixed thoroughly and measured via $\gamma$-spectroscopy. The $\gamma$-spectroscopy was performed with a 151\,cm$^3$ HPGe detector from Canberra Industries (1.8\,keV resolution at the 1332\,keV $^{60}$Co peak; 50.1\,\% relative efficiency), connected to a PC-based multi-channel analyzer with preloaded filter and Loss-Free Counting (LFC) system.

The powder was then combined with a previous batch of $^{229}$ThF$_4$:PbF$_2$:CaF$_2$ (15\,mg, containing 0.3\,MBq $^{229}$Th) to give a total amount of 45\,mg $^{229}$ThF$_4$:PbF$_2$:CaF$_2$ with a weight ratio of 0.33:1:14 and a total activity of 5.8\,MBq $^{229}$Th used. The previous batch originated from a trial run of above mentioned method. The stock powder was then split in 3 parts of equal amounts by weight (15\,mg) to give three aluminum containers with equivalent material (No. 1, No. 2, No. 3). All vials were measured with $\gamma$-spectroscopy in the same geometry to check the distribution of activity between the containers (0.33:1:14, as above) and stored in a desiccator until used. Due to losses during the process 4.7\,MBq out of 5.8\,MBq $^{229}$Th were obtained as usable powder.

\subsection{Preparation of \textsuperscript{232}ThF\textsubscript{4}:PbF\textsubscript{2}:CaF\textsubscript{2} growth material}

For calibration purposes, process optimization, and many measurements, which do not probe nuclear properties (e.g. VUV transmission measurements), commercially available $^{232}$Th can be used as a proxy. 

Two different methods for the $^{232}$ThF$_4$:PbF$_2$:CaF$_2$ sample preparation were used. Method A was used to prepare \textsuperscript{232}Th containing powder analogous to the \textsuperscript{229}Th powder. Then conditions (ratios between components, water content, contamination, etc.) were kept constant between growing \textsuperscript{232}Th:CaF\textsubscript{2} and growing \textsuperscript{229}Th:CaF\textsubscript{2}. Method B was used when different ratios between components should be tested. This approach avoided the lengthy coprecipitation which allowed for quick testing. Losses of Th were not important in this method due to the use of the abundant \textsuperscript{232}Th isotope. The two methods were: \par

\begin{itemize}
    \item[A:] The sample was prepared analogous to the $^{229}$ThF$_4$:PbF$_2$:CaF$_2$ procedure described above using $^{232}$ThNO$_3$ instead of $^{229}$ThNO\textsubscript{3}. 
    A small amount of $^{229}$ThF$_4$:PbF$_2$:CaF$_2$ was added (1\,kBq of $^{229}$Th) to the $^{232}$ThF$_4$:PbF$_2$:CaF$_2$ powder for tracing purposes.
    \item[B:] The sample was prepared by mixing commercially available CaF$_2$, PbF$_2$ and $^{232}$ThF$_4$. Then 1\,kBq of $^{229}$ThF$_4$ was added by letting a spoon touch first the produced $^{229}$ThF$_4$ then the $^{232}$ThF$_4$ for tracing the location of the dopant after growth.
\end{itemize}

\subsection{Vertical gradient freeze growth process}
To grow 3.2\,mm diameter, 11\,mm long crystals, we use a modified vertical gradient freeze method. It was first developed in 1924 by Stöber~\cite{Stober1925}, inspired on the Bridgman method. This method was adapted to our needs in cooperation with the Fraunhofer institute for integrated systems and device technology (IISB) to grow crystals with minimal dopant losses~\cite{schreitl2016growth}. The main advantage of the VGF method over other methods like Czochralski is that the growth speed is decoupled from the crystal diameter. Very small diameter crystals and hence high (10$^{18}$\,cm$^{-3}$) doping concentrations can be realized with the extremely scarce $^{229}$Th isotope while maintaining low growth speed ($<$0.5\,mm/h) which promotes high quality crystals and improved VUV transparency.

\begin{figure}[h!]
\includegraphics[width=\linewidth]{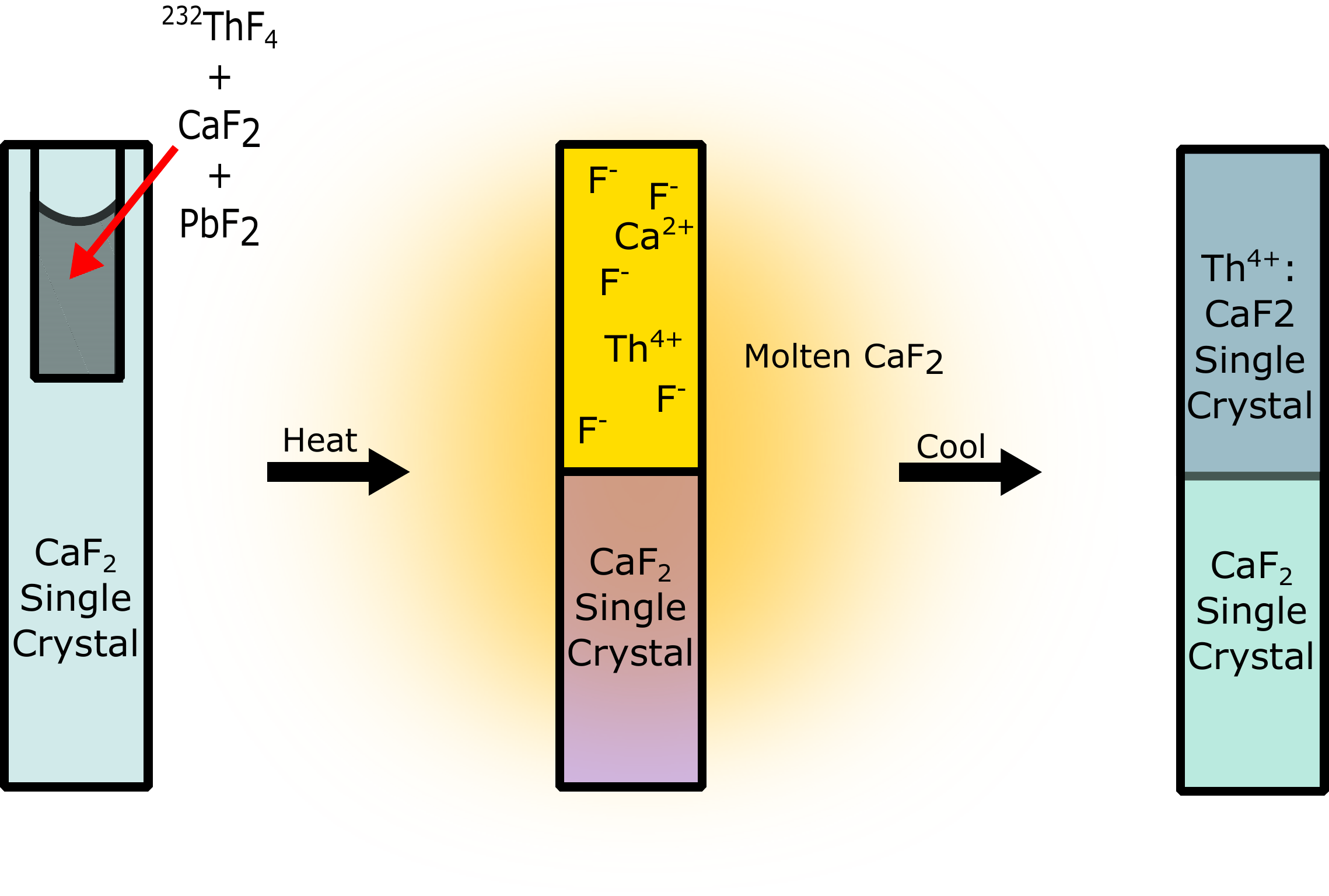}
\caption{Simplified schematic representation of the vertical gradient freeze method applied to the seed crystal in figure~\ref{fig:seedcrystal} filled with dopant powder. While applying a steep temperature gradient, the top of the crystal melts and becomes a liquid while the bottom remains solid. By slowly cooling and moving the melting interface upwards, a doped single crystal can be grown on the seed crystal.}
\label{fig:growthproc}
\end{figure}

\begin{figure}[h!]
\includegraphics[width=\linewidth]{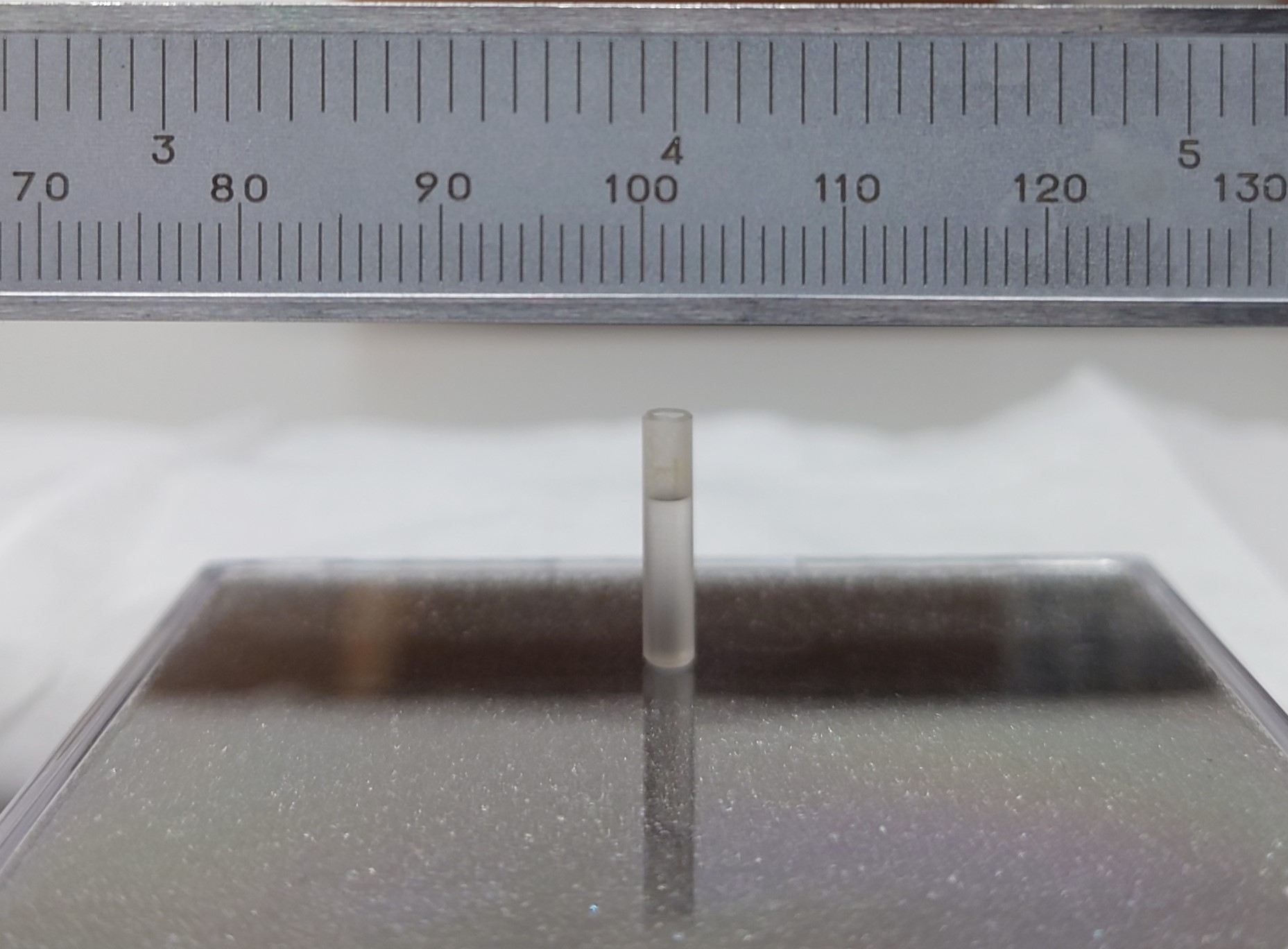}
\caption{Seed crystals with 3.2\,mm diameter, 11\,mm length and a 2\,mm diameter, 5\,mm deep pocket used for growing highly doped \textsuperscript{229}Th:CaF\textsubscript{2}. A caliper with inch and mm scales is shown as well.
}
\label{fig:seedcrystal}
\end{figure}

\begin{figure*}[ht]
\includegraphics[width=\linewidth]{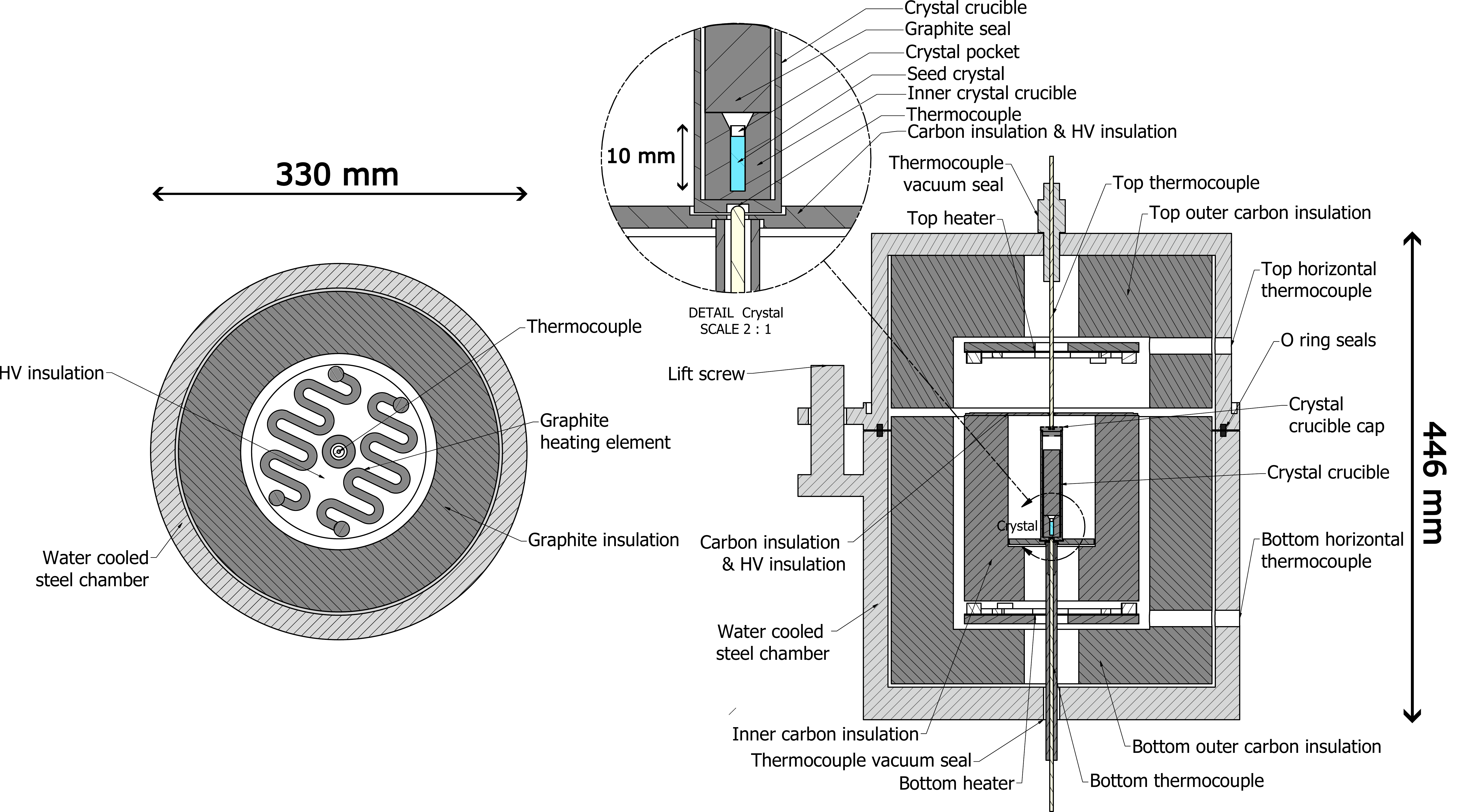}
\caption{Crystal growing device used to grow 3.2 mm diameter crystals. On the left side is a horizontal cut displaying the graphite heating element with the isolated thermocouple through the center. The right image displays a vertical cut that shows all of the main components of the vertical gradient freeze growing machine and zooms into the crystal location. The thermocouples are platinum/rhodium (Pt30Rh-Pt6Rh) with an alsint casing. The vacuum seals for the thermocouples are made of FKM, which is able to withstand both the low pressure fluorine atmosphere and the high temperatures which are present at the ends of the thermocouple. Water is circulated through the steel vacuum chamber to provide cooling.}
\label{grower}
\end{figure*}

\begin{figure}[h!]
    \centering
    \includegraphics[width=\linewidth]{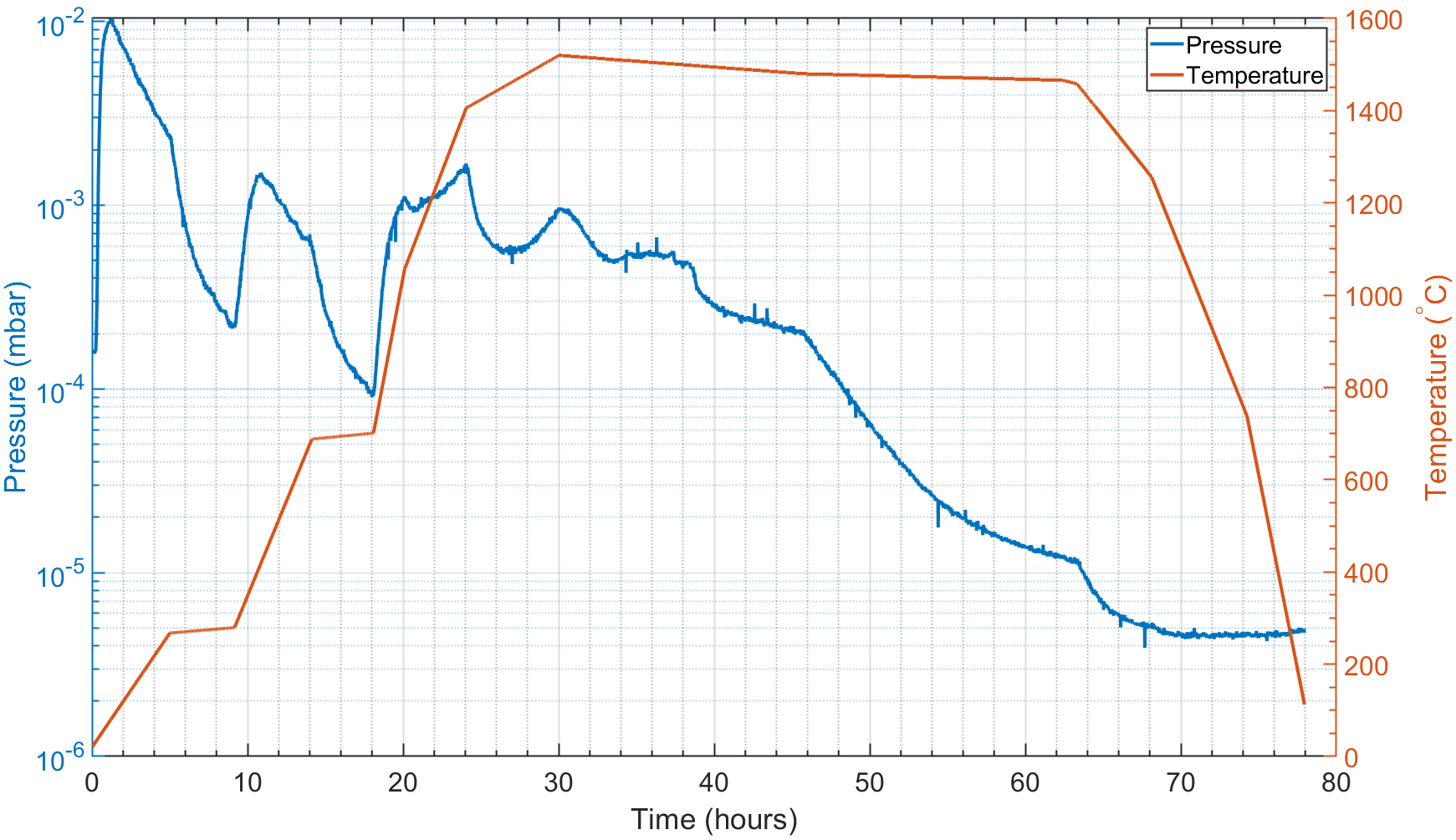}
    \caption{Pressure evolution during the temperature cycle versus time to grow 3.2\,mm diameter crystals. Temperature was taken from the top vertical thermocouple. One can see that every increase in temperature increases the pressure, which after temperature stabilization decreases again. Interestingly the pressure does not significantly increase anymore at temperatures above 1000\,\textcelsius{}.}
    \label{fig:growP}
\end{figure}

In the vertical gradient freeze method, a temperature gradient is slowly driven across the starting growth material (powder and seed) to control the liquid-solid interface, as illustrated in figure~\ref{fig:growthproc}. In our implementation, sub-millimeter control of this interface layer is required, imposed by the small crystal dimensions (see figure~\ref{fig:seedcrystal}). The gradient is realized by electronically controlling the currents in two ohmic heaters on top of and below the crucible containing powder and seed (see figure~\ref{grower}). The entire system is kept under vacuum during the growth process, approximately 10$^{-4}$\,mbar at the start of the growth (see figure~\ref{fig:growP}). The vacuum prevents oxidation of the graphite and CaF$_2$ powder. Inert gases could also be used for this purpose, however achieving a steep temperature gradient (20\,K/cm) is then very challenging.

 A stationary graphite crucible is filled with a CaF$_2$ seed crystal together with the starting material in the seed pocket. A temperature gradient of $\approx$\,20\,\textcelsius{}/cm should be around the melting temperature of the chosen material, which is 1418\,\textcelsius{} for CaF$_{2}$. In this way, the powder in the pocket of the seed crystal can be molten together with the top part of the seed. Due to the rather steep gradient the bottom of the seed does not melt. The freezing interface is then slowly moved upward such that the melt can crystallize on top of the seed crystal, and thus grow a single crystal following the orientation of the seed.

The diameter and length of the crystals grown in our laboratory has been successively reduced from 17\,mm to 3.2\,mm and from 40\,mm to 11\,mm, respectively, by careful calibration and using smaller seeds. This implements a volume reduction and corresponding concentration increase by a factor $\approx$100. Seed crystals of 5\,mm diameter (with different, single crystal orientations) were purchased from Korth, Matek, and Alkor. Hyperion Optics milled the seeds down from 5\,mm to 3.2\,mm diameter and drilled a 2\,mm diameter hole (5\,mm deep) into the top, to accommodate the doping material. Figure~\ref{fig:seedcrystal} shows a photograph of such a prepared seed, and the DETAIL in figure~\ref{grower} shows the positioning of the seed inside the graphite crucible.

The pocket is filled with the co-precipitated growth material. Special care was taken to use metallic funnels just as the containers to avoid material losses due to electrostatic adsorption to the wall. The pocket within the seed ensures minimal losses and easy handling of the crystals before starting the VGF process. The filled seed is placed into the VGF furnace (figure~\ref{grower}) inside of a carbon crucible. The phase diagram of CaF$_2$ and ThF$_4$ at low vacuum has not been measured to our knowledge, but we observe evaporation. 
Fluorine was detected by using a Pfeiffer quadrupole mass spectrometer and traces of radioactive Th and U in the graphite show that these partly evaporate during growth. The volume above the seed in which material can be evaporated is small in our setup. We observed that this small volume increases doping efficiency as opposed to a larger volume or direct connection to the vacuum pump. We hypothesize that when the vapor is actively pumped away more of the material is evaporated which reduces doping concentration.

The seed is then grown to a single crystal: Two ohmic carbon heaters (see left side of figure~\ref{grower}) are used to create the steep temperature gradient over the crystal, partially melting it. Short-term temperature stability is maintained with short horizontal alsint-insulated thermocouples close to the heaters and absolute calibration is done with long vertical thermocouples close to the crucible. 

The temperature at the crystal pocket is not exactly the same as the one indicated by the bottom thermocouple, thus we employ an iterative temperature calibration process to ensure that the temperature is such that the crystal only melts midway. Important here is to carefully monitor the in and outside temperature of the thermocouples to get an accurate temperature measurement and avoid drifts in the system (e.g. by room temperature or cooling water). The graphite thermal isolation and alsint isolation of the thermocouples must not touch, since at high temperatures these two chemically react, which slowly degrades the casing of the thermocouples which can create vacuum leaks.

\begin{figure}[h!]
    \centering
    \includegraphics[width=\linewidth]{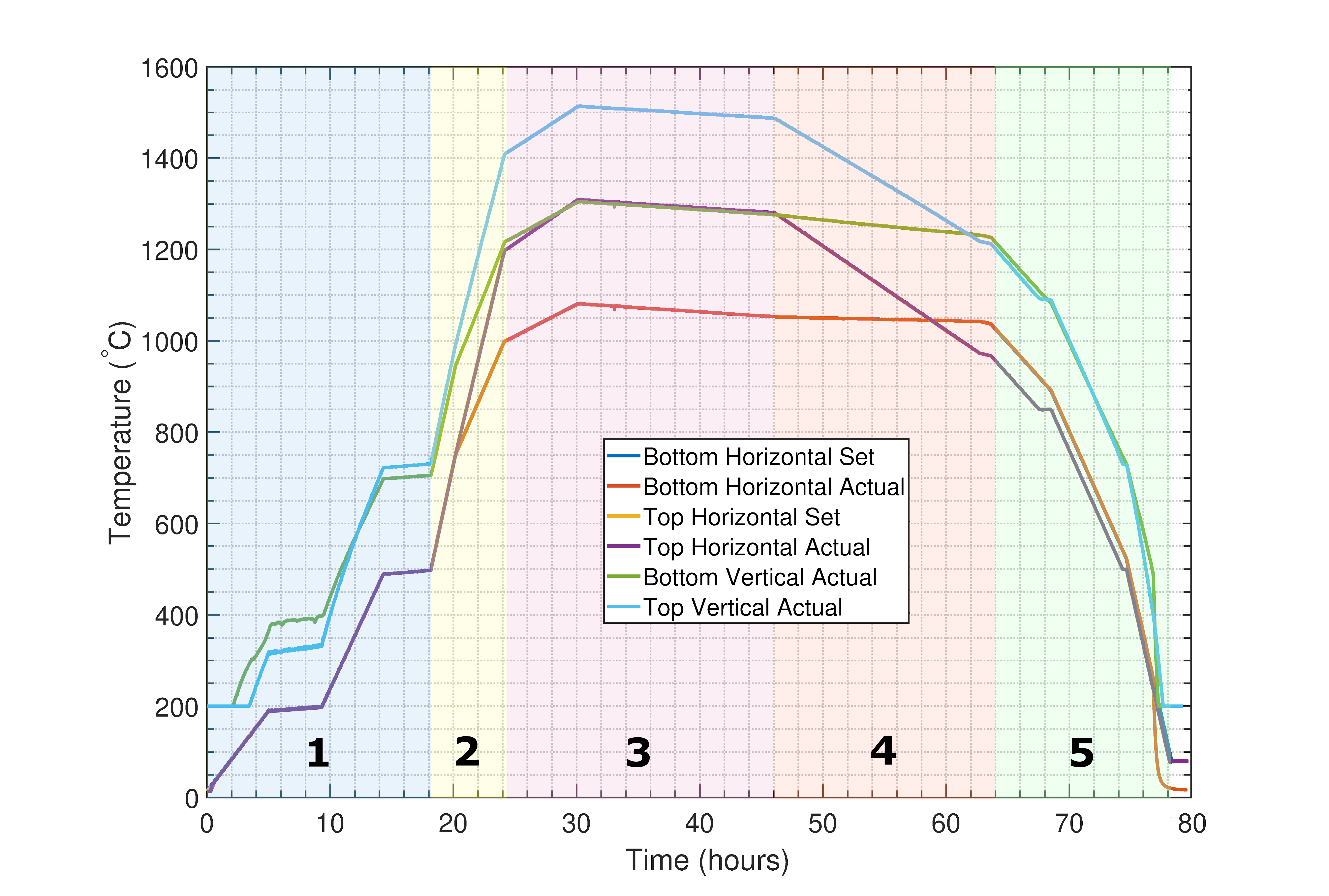}
    \caption{Temperature cycle versus time to grow 3.2\,mm diameter crystals as set and measured by the thermocouples indicated in figure~\ref{grower}. The sections as described in the text are indicated as well. The entire process takes about 80~hours or approximately 3~days. Note that only the horizontal thermocouples are used to actively control the temperature, and thus only these have set temperatures. Their temperatures are lower than the vertical thermocouples as they are further from the center of the crucible. The lines of the set temperatures are almost completely hidden by the actual measured lines.}
    \label{fig:growT}
\end{figure}

The temperature cycle of the growing process (shown in figure~\ref{fig:growT}) is divided into five sections: 1) 18~hours of heating up the system, outgassing, and restoring pressure (see figure~\ref{fig:growP}) 2) 6~hours of scavenging oxygen through reaction with PbF$_2$ 3) 22~hours of melting the top half of the crystal and also slowly freezing it 4) 18~hours of annealing the crystal 5) 14~hours of cooling down. A vacuum of at least 10$^{-4}$\,mbar is obtained before growth. During growth (especially during section 1) the pressure can go up to 10$^{-2}$\,mbar. The complete growth process typically takes 3~days. \par

Due to the radioactive nature of the dopant, special security measures are implemented in the growth process. On the pre-pump, carbon filters are installed to absorb any evaporated material. The dopant material also is absorbed into the graphite insulation, which becomes radioactive after several growth cycles. In every growth, the insulation absorbs and also releases some dopant; cross-contamination of dopants was observed in a pure CaF\textsubscript{2} crystal, grown after growth of a radioactively doped crystal. The growth of undoped crystals can be used to reduce cross-contaminations by absorbing them, we observe a reduction of roughly a factor 10 per growth process.

One important aspect in growing CaF$_2$ crystals is the probability of incorporating oxygen, especially at higher temperatures~\cite{molchanov2005study,ko2001czochralski}. Oxygen contamination is known to reduce the transparency of CaF$_2$, especially in the VUV region. At elevated temperatures, the carbon crucible should react with any background gaseous O$_2$ and H\textsubscript{2}O to form CO in the system. The H$_2$O that is adsorbed on the surface of the crystal powder, however, will first react with the CaF\textsubscript{2}. The H\textsubscript{2}O will react with the CaF$_2$ to CaO (T\textsubscript{melt}=2613\,\textcelsius{}~\cite{haynes2016crc}) and HF according to

\begin{equation}
    \text{CaF}_2^{s} + \text{H}_2\text{O}^{a} \xrightarrow{} \text{CaO}^{s} + 2\text{HF}^{g}.
    \label{hydration}
\end{equation}

To mitigate this, oxygen scavengers are used: Fluoride compounds (PbF\textsubscript{2} T\textsubscript{melt}=830\,\textcelsius{}~\cite{haynes2016crc}) in the powder that before melting CaF\textsubscript{2} react preferentially with oxygen and water to volatile oxygen containing compounds (PbO T\textsubscript{melt}=887\,\textcelsius{}~\cite{haynes2016crc}) which are evaporated at higher temperatures and transported away such that the oxygen is replaced by fluoride:

\begin{equation}
    \text{CaO}^{s} + \text{PbF}_2^{g} \xrightarrow{} \text{CaF}_2^{s} + \text{PbO}^{g},
    \label{fluoridation}
\end{equation}
\begin{equation}
    \text{H}_2\text{O}^{a} + \text{PbF}_2^{g} \xrightarrow{} 2\text{HF}^{g} + \text{PbO}^{g}.
    \label{hydroxidation}
\end{equation}

The gaseous lead(II) oxide will then react with cooler carbon in the insulation walls where they form metallic lead (T\textsubscript{melt}=328\,\textcelsius{}~\cite{haynes2016crc}) and CO (not CO\textsubscript{2} according to the Boudouard equilibrium)

\begin{equation}
    \text{PbO}^{g} + \text{C}^{s} \xrightarrow{} \text{Pb}^{s} + \text{CO}^{g}.
    \label{carbonation}
\end{equation}

\subsection{Temperature calibration}
Melting a mixture of CaF$_2$ and ThF$_4$ will lead to a modified melting temperature compared to the separate constituents, as investigated in~\cite{capelli2015thermodynamic}. Since the amount of available $^{229}$Th will not allow growing doping concentrations higher than 0.05\,wt\,\% (1\,MBq) to 0.5\,wt\,\% (10\,MBq) we do not expect (and did not observe) significant changes in the melting temperatures. The temperature calibration could be used independent of the doping concentration, and a partial phase transition into CaThF$_6$ occurring at 19\,wt\,\% can be ignored.

\begin{figure}
\includegraphics[width=\linewidth]{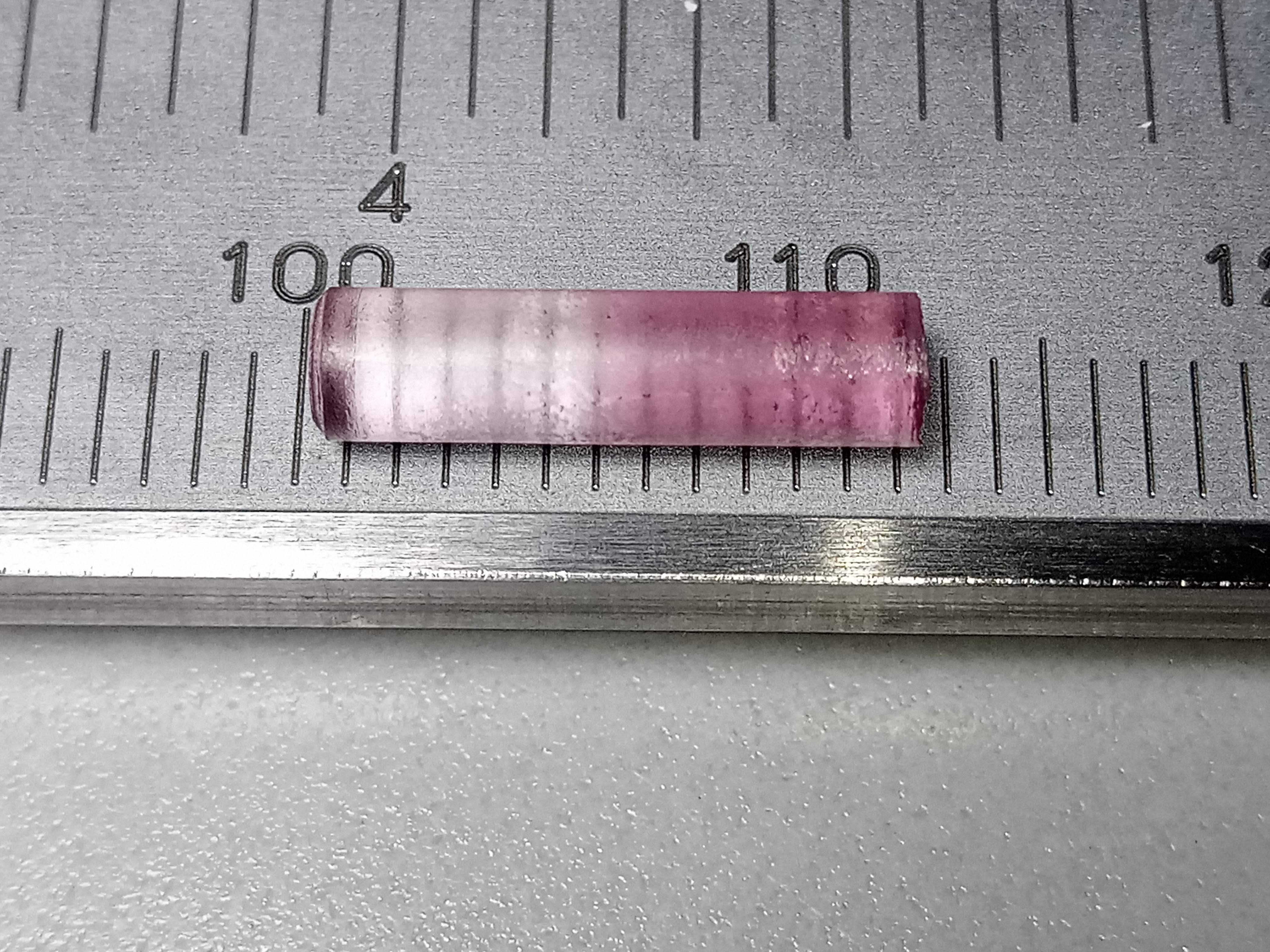}
\caption{$^{238}$U:CaF$_2$ single crystal grown during temperature calibration. The crystal lies on a caliper where the mm scale is indicated below. The melting boundary can be clearly seen in the color difference. The bottom (left) part is only slightly colored due to Uranium outgassing from the carbon crucible that accumulated there in earlier growth cycles and penetrating the outer layer of the CaF$_2$ crystal. The top of the crystal on the right is strongly colored due to the incorporation of the dopant material during melting and re-solidification.}
\label{udoped}
\end{figure}

To finely adjust the temperature settings of the growth process, low concentrations of uranium-238 were used as a colorant. As shown in figure~\ref{udoped}, the coloring allows to accurately determine the lowest position of the liquid-solid interface, in other words, how deeply the melting progressed into the seed crystal. For a typical gradient of 20\,\textcelsius{}/cm and a peak (set) offset temperature of 1418\,\textcelsius{}, the precision (reproducibility) of the melting boundary is $~$1\,mm with a change in 2.5\,\textcelsius{} in set temperature shifting by 1\,mm. The calibration can be transferred from uranium to thorium without adaptation due to the low doping concentrations. The calibration is obtained in an iterative process involving 5-10 growth processes. The temperature interval between obtaining a fully molten crystal and not melting the seed at all is about 15\,K, the temperature relation is not fully linear and the drilled top part of the seed crystal containing the growth material is more susceptible to melting than the lower bulk of the seed crystal.

Crystals grown with $^{229}$Th and $^{232}$Th doping are optically fully transparent, so it is less obvious to determine the melting depths. Here, the radioactivity of $^{229}$Th can be used to extract spatial information on the doping distribution (see below). For doping with $^{232}$Th, which is used as a proxy in experiments not focusing on nuclear properties, the growth material is "spiked" with a low (few Bq) activity of $^{229}$Th to produce a measurable signal on a $\gamma$-detector. Both thorium isotopes are chemically identical and therefore distribute identically during the growth process.

\subsection{Doping efficiency and homogeneity}

Before and after growth, $^{229}$Th-doped crystals are measured on a $\gamma$-spectrometer in a reproducible geometry (in the crucible) to determine the efficiency of the doping process by comparing the intensity of the 193 keV $\gamma$-line. We observe a 20-30\% doping efficiency of the starting material where we attribute the losses to evaporation. The highest doping concentration realized so-far was with a $^{229}$Th activity of 1\,MBq (3.6$\cdot10^{17}$ nuclei) in the starting material, reaching a concentration of 2.4$\cdot10^{17}$\,cm$^{-3}$ (see figure~\ref{fig:229vuv}).

\begin{figure}
    \centering
    \includegraphics[width=\linewidth]{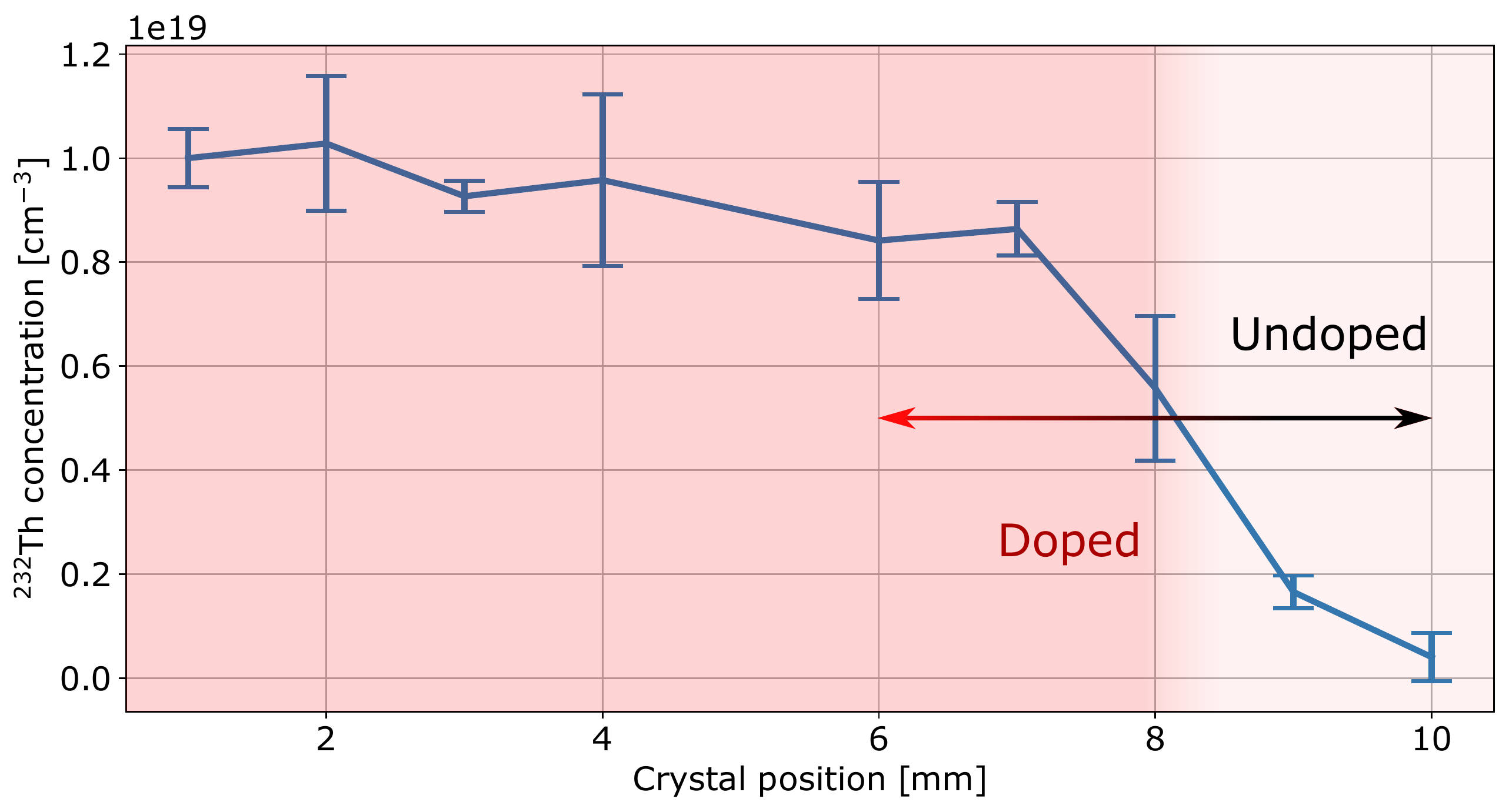}
    \caption{Doping concentration versus position in a ("spiked") \textsuperscript{232}Th:CaF\textsubscript{2} crystal. The red color indicates the doped and undoped parts, although the thorium is colorless. The grown crystal was cut in 1 mm pieces and the activity of each piece was determined with the $\gamma$-detector and related to concentration through a reference sample. In the figure, 0 mm corresponds to the top and thus doped part of the crystal. Error bars indicate the uncertainty in the obtained count rate in the $\gamma$ detector.}
    \label{fig:c5}
\end{figure}

The doping concentration is homogeneous over the molten part of the crystal and decays over 2-3 mm in the interface to the not molten seed, as illustrated in figure~\ref{fig:c5}. For this investigation, a crystal doped with $^{232}$Th, spiked with $^{229}$Th, was grown. After growth the crystal was cut into 1\,mm disks and each disk was measured in a reproducible geometry in the gamma-detector. Again, the 193\,keV line activity was detected and from this the concentration was calculated by measuring a reference sample. The seed crystal was molten to a depth of around 6\,mm, in which the concentration is approximately homogeneous (within 2$\sigma$ as seen in figure~\ref{fig:c5}). Beyond, we observe diffusion of dopant into the not molten part on a length scale of about 2\,mm, until no doping can be detected in the not molten region of the seed crystal.


\begin{figure}
\includegraphics[width=\linewidth]{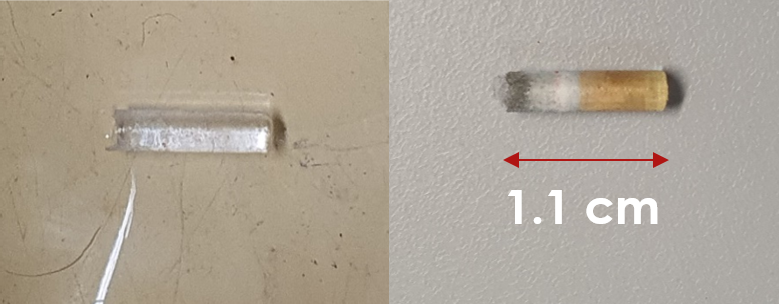}
\caption{$^{229}$Th:CaF$_2$ single crystal grown with an activity of 1\,MBq in the growth material. On the left, the crystal as it was inspected immediately after growth. On the right, the crystal after 3\,days. The orange coloring is due to the agglomerations of defects, F-centers, produced by the radioactivity~\cite{beeks2022nuclear}. The melting boundary can be clearly seen in the color difference. The top of the crystal is on the right.}
\label{thdoped}
\end{figure}

\subsection{VUV transmission}

Good transmission of the obtained crystals in the VUV range around 150\,nm is imperative for all attempts to optically manipulate (or detect) the $^{229}$Th isomeric state. We therefore performed a series of characterization experiments described below, with a special emphasis on the behavior of the VUV absorption with doping concentration.

Doping Th into the CaF$_2$ matrix should not reduce the bandgap and therefore the optical transmission window significantly, as theoretically calculated~\cite{dessovic2014229thorium} and measured~\cite{schreitl2016growth}. However, additional electronic states (often referred to as color centers or defect states) can emerge within the bandgap~\cite{Nickerson2020color}. 

In order to measure the VUV transmission, the crystals were cut and polished. Cutting was done using a Wiretec DWS100 with a 0.08\,mm diamond coated wire. Facets cut with the wire saw were flat enough to be polished in a 1-step process. Polishing was done with a Buehler crystal polishing machine and a Buehler SiC P4000 grit Silicon Carbide polishing paper. The polishing paper was wetted with isopropanol. As CaF2 is hygroscopic and adsorbed water decreases transmisson, the polishing paper was wetted with isopropanol instead of water~\cite{reiterov1980influence}.

\begin{figure}[h!]
    \centering
    \includegraphics[width=0.6\linewidth]{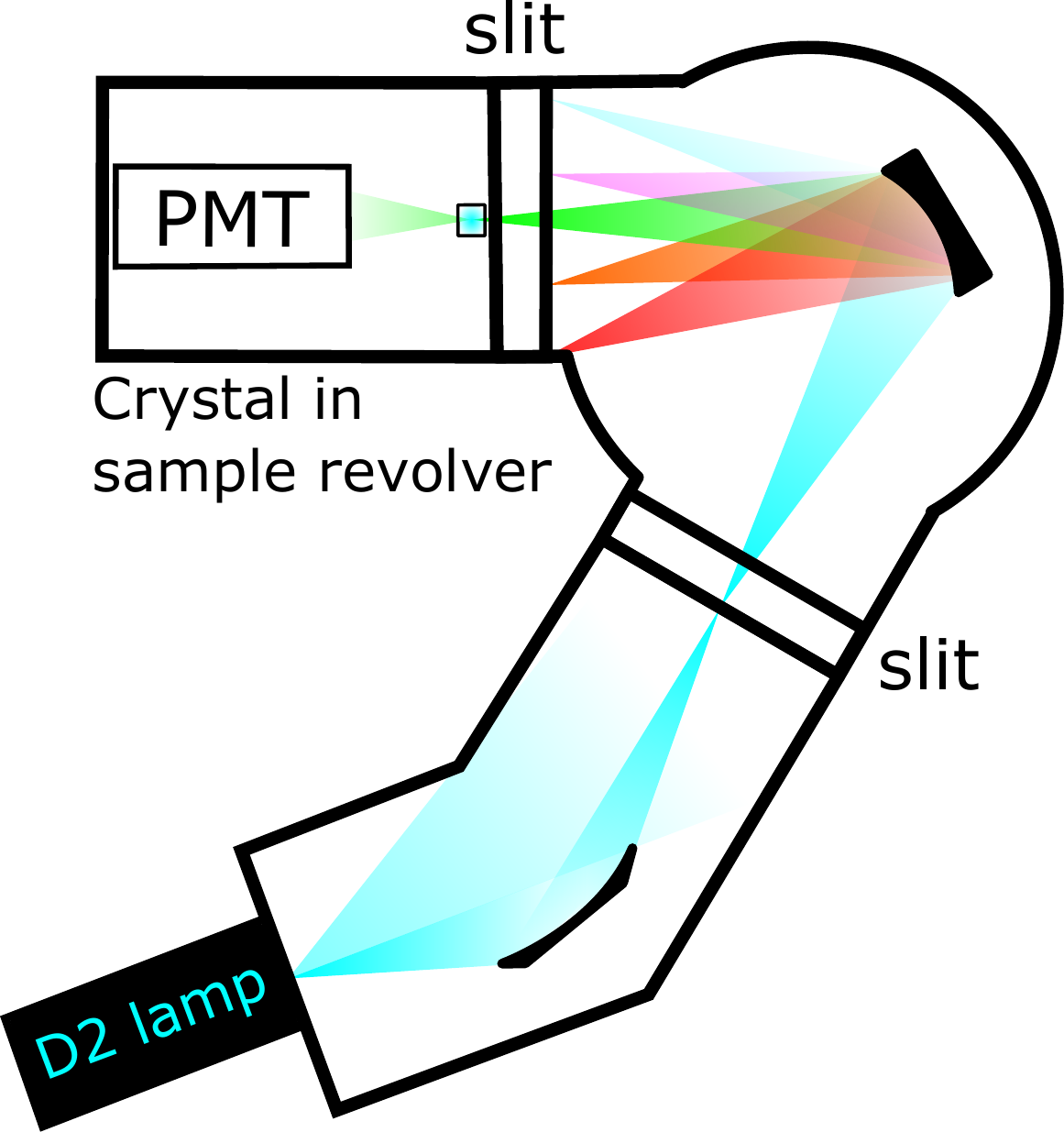}
    \caption{Experimental setup to measure the wavelength-dependent transmission of a crystal. The light produced by a deuterium (D2) VUV lamp is focused by a mirror onto a monochromator which, using a rotating focusing reflective grating, can image different colors of light through a sample in a filter wheel on a PMT detector.  By comparing the light output at a set wavelength with and without crystal in the optical path, a transmission spectrum can be measured.}
    \label{fig:transpmt}
\end{figure}

Transmission measurements were performed using a dedicated setup as schematically depicted in figure~\ref{fig:transpmt}. The light of a Hamamatsu L15094 D\textsubscript{2} lamp is focused with a toroidal mirror onto the entrance slit of a McPherson 234/302 monochromator. Part of the light does not hit the focusing mirror. The light is separated into its spectral components by the grating and is focused onto the exit slit. By rotating the grating the exit wavelength can be selected. The exit slit cuts out a small portion of the spectrum effectively creating a narrow wavelength light source with a linewidth down to 0.1\,nm. The linewidth can be changed by changing the entrance/exit slit width (0.01\,mm to 2.50\,mm). This light travels through the crystal, and is recorded by a Hamamatsu R6835 head-on CsI photomultiplier tube (PMT) which is mounted close to the crystal.

Although conceptually simple, measuring a wavelength-dependent absolute absorption is burdened with several experimental challenges. These are connected with geometrical changes in the beam paths due to the presence of the sample (beam shifts and astigmatism), strong spectral intensity modulations and overall intensity instabilities in the VUV source (deuterium lamp). These lead to an overall systematic error on the following transmission measurements of $\pm$5\,\%.


\begin{figure}[h!]
    \centering
    \includegraphics[width=\linewidth]{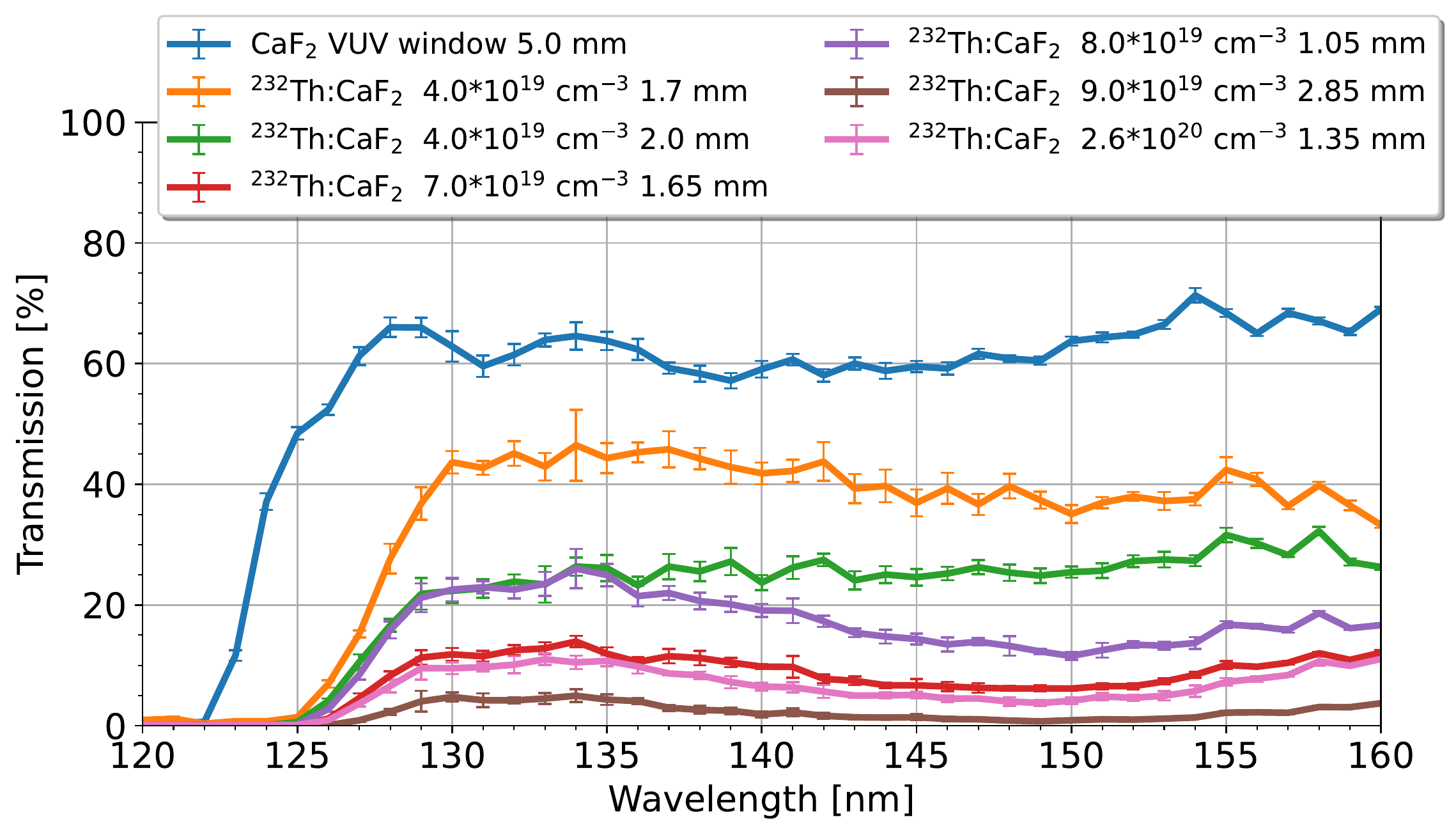}
    \caption{Spectral transmissions of \textsuperscript{232}Th:CaF\textsubscript{2} with different thicknesses and concentrations for different wavelengths. A sample of VUV grade CaF\textsubscript{2} produced and polished by Korth GmbH is displayed for comparison, thickness 5\,mm. It can be seen that the $^{232}$Th doped crystals all absorb heavily below 130\,nm, but pure CaF\textsubscript{2} only below 125\,nm. Due to the different thicknesses of the samples the transmission might be different, but the absorption coefficient not.}
    \label{fig:thvuv}
\end{figure}

\begin{figure}[h!]
    \centering
    \includegraphics[width=\linewidth]{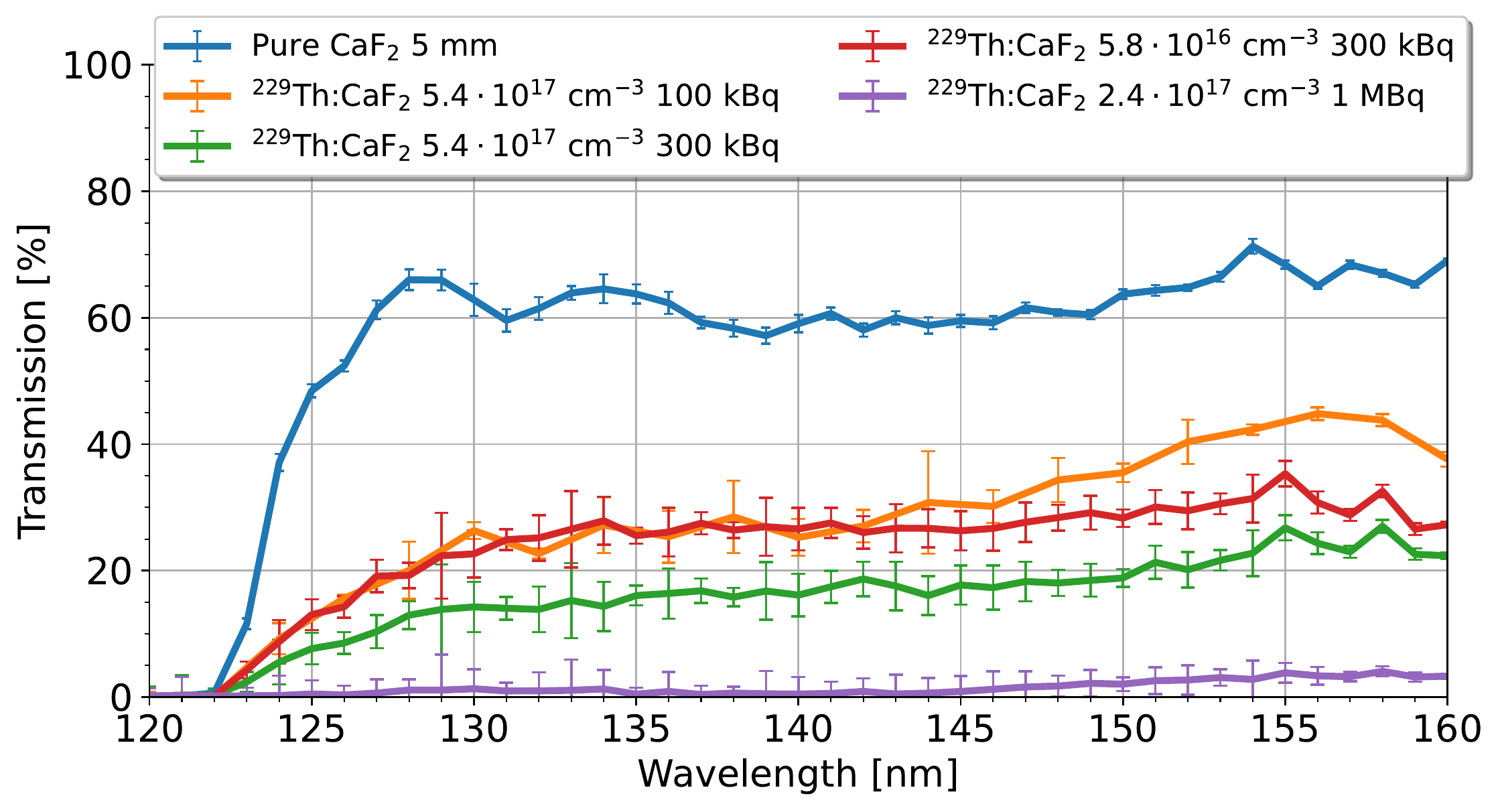}
    \caption{Spectral transmissions of \textsuperscript{229}Th:CaF\textsubscript{2} with different thicknesses and concentrations for different wavelengths. A sample of VUV grade CaF\textsubscript{2} produced and polished by Korth GmbH is displayed for comparison (not remeasured), thickness 5\,mm. It can be seen that the $^{229}$Th doped crystals are consistently less transparent than pure CaF\textsubscript{2}, but otherwise follow the same behavior.}
    \label{fig:229vuv}
\end{figure}

\begin{figure}[h!]
    \centering
    \includegraphics[width=\linewidth]{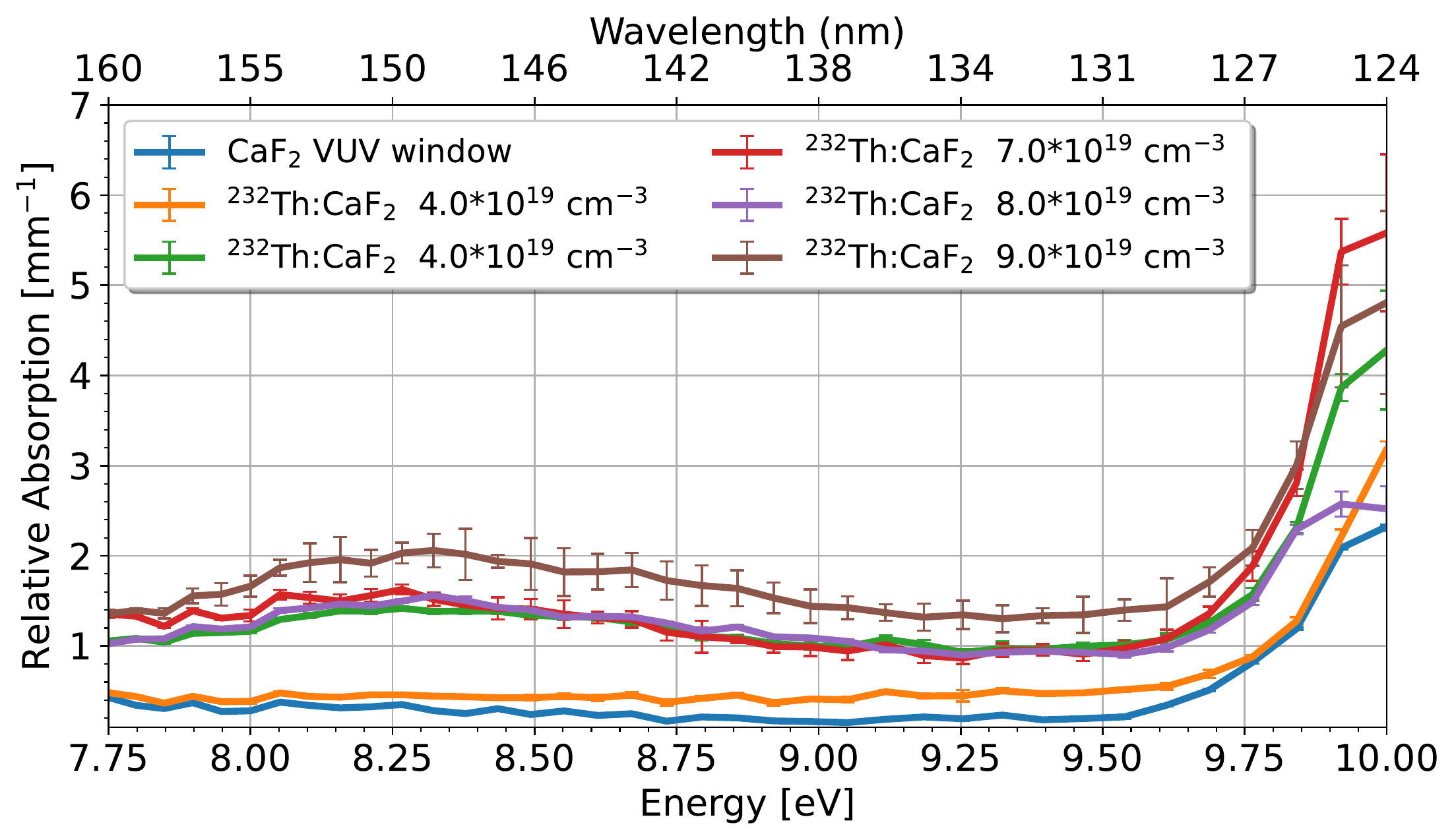}
    \caption{Relative absorption coefficient of \textsuperscript{232}Th:CaF\textsubscript{2} with different concentrations for different energies and wavelengths. Main features are general increase in absorption for higher concentrations, absorption around 150\,nm and below 130\,nm or after 9.5\,eV. The thickness of the crystal limits how well high absorption can be measured which is an artefact in the figure: the thinnest crystals have the highest absorption around 10\,eV.}
    \label{fig:thabs}
\end{figure}

\begin{figure}[h!]
    \centering
    \includegraphics[width=\linewidth]{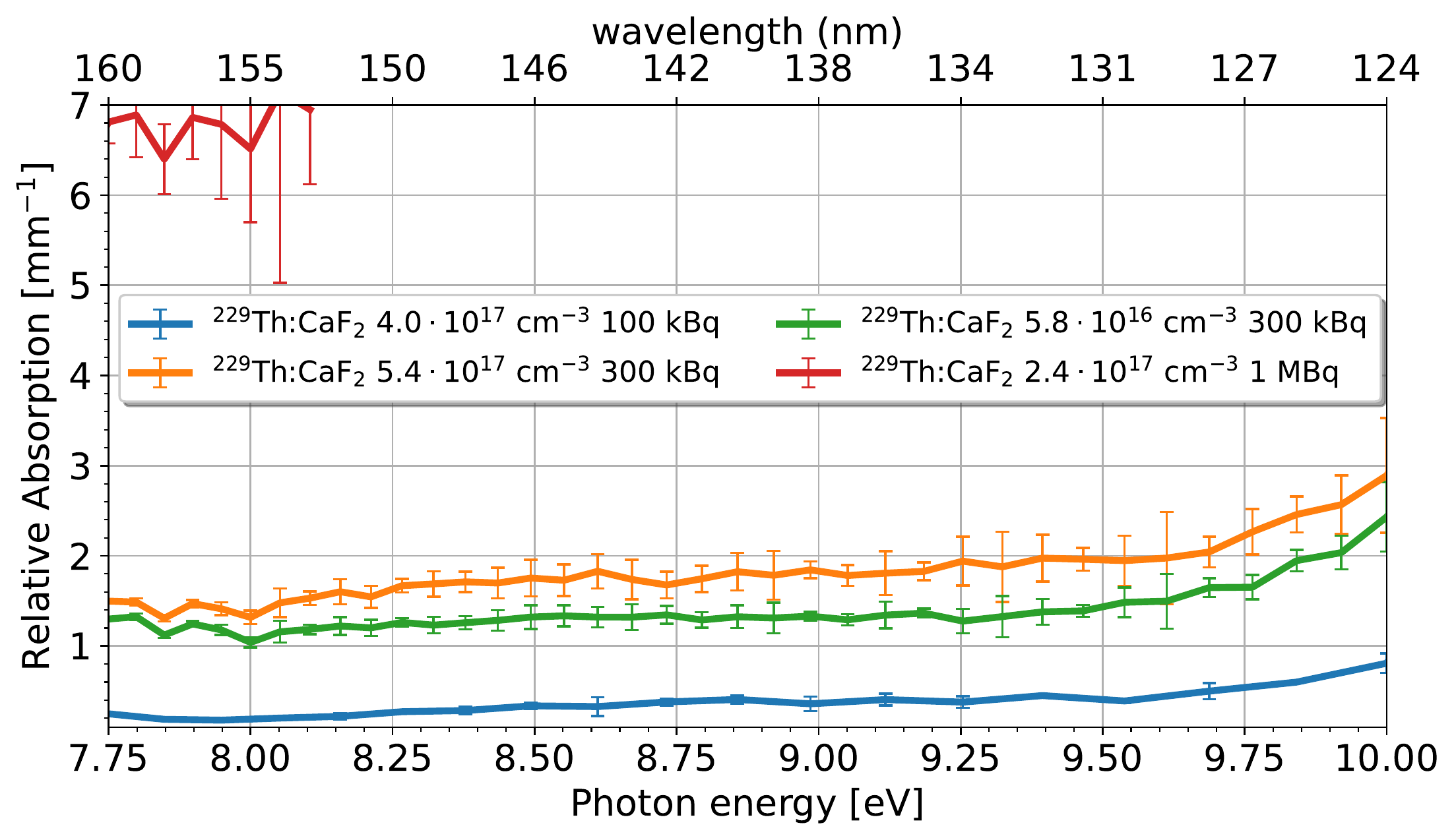}
    \caption{Relative absorption coefficient of \textsuperscript{229}Th:CaF\textsubscript{2} crystals with different concentrations for different energies and wavelengths. The crystal grown with 1 MBq activity, was only 0.5 mm thick. The thickness allowed an absorption measurement of only the high wavelength region. The absorption profile of these crystals is flat with few features. The bandgap edge of CaF\textsubscript{2} after 10\,eV is seen. With higher activity during growth, more absorption is observed independent of concentration.}
    \label{fig:229abs}
\end{figure}

To compare measurement results between different samples, we account for the thickness of the crystal, surface reflections, and absorption at the surface by normalizing the absorption coefficient to pure (undoped) CaF\textsubscript{2}. This relative absorption coefficient $\mu_{rel}$ then is a measure for the absorption caused by Th doping in the bulk and at the surface. 
Although generally relevant in CaF\textsubscript{2}, we expect two-photon absorption processes to be negligible in these measurements, due to the very low intensities used. It is also assumed that surface quality is similar for all measured crystals. However, this is hard to guarantee due to the hygroscopic nature of calcium fluoride, which leads to water adsorption~\cite{denks2000dependence} 

The relative absorption coefficient is defined as 
\begin{equation}
    \mu_{rel} = \frac{-\log{(T/T_{CaF_2})}}{d},
\end{equation}
where $T$ is the transmission, $T_{CaF_{2}}$ the transmission of pure CaF\textsubscript{2} and $d$ the thickness of the crystal. We measure the transmission and subsequently calculate the relative absorption of \textsuperscript{229/232}Th:CaF\textsubscript{2}, for the wavelength region of interest 120-160\,nm. 

Non-normalized transmission measurements for \textsuperscript{232}Th- and \textsuperscript{229}Th-doped crystals can be seen in figures~\ref{fig:thvuv} and~\ref{fig:229vuv}, respectively. Figures~\ref{fig:thabs} and~\ref{fig:229abs} show the corresponding relative absorption coefficients. 


From figures~\ref{fig:thvuv} and \ref{fig:229vuv} it can clearly be seen that all grown single crystals have a transmission above 1\,\% around 150\,nm, some reaching $\approx$\,40\,\%. In cases where the seed crystal was fully molten during the growth process, we obtain completely VUV-opaque samples (transmission $<0.1\,\%$, independent of doping concentration). We conjecture that in these cases, a polycrystal is formed, which suppresses VUV transmission due to the presence of grain boundaries.

Doping concentrations up to 2.6$\cdot$10\textsuperscript{20}\,cm\textsuperscript{-3} for \textsuperscript{232}Th and up to 5.4$\cdot$10\textsuperscript{17}\,cm\textsuperscript{-3} \textsuperscript{229}Th were grown with good transparency. \par

The \textsuperscript{232}Th doped crystals often show an absorption around 150\,nm, which we attribute to Ca metallic particles~\cite{rix2011formation}. In undoped CaF$_2$ they absorb around 160\,nm but the presence of Th changes the refractive index of the crystal, thereby changing the absorption wavelength of these centers. Higher doping concentrations generally seem to lead to more absorption.

In~\cite{Angervaks:18}, sharp absorption bands around 160-170\,nm were observed and attributed to calcium metallic precipitates or colloids incorporated in the CaF$_2$ matrix. These precipitates can form due to crystal damage or a deficiency of fluoride.

All grown Th:CaF$_2$ crystals seem to absorb starting from 130\,nm and have very low transmission below 125\,nm, earlier than the transmission edge of CaF\textsubscript{2} which starts at 125\,nm and has little transmission below 122\,nm. \par

In the absorption of \textsuperscript{232}Th (figure~\ref{fig:thabs}) it can be seen that crystals with similar concentrations have similar absorption. Three features can again be identified: Increasing absorption around 150\,nm or 8.2\,eV, a general increase of absorption with increasing doping concentration and very strong absorption starting at $<$130\,nm or 9.5\,eV. These features are all compared to pure CaF\textsubscript{2}, so they are an effect of the dopant. \par


In figure~\ref{fig:229vuv} it can be seen that the \textsuperscript{229}Th doped crystals are transparent up until the transmission edge of undoped CaF\textsubscript{2}. The overall transmission of these crystals is lower. In the absorption of \textsuperscript{229}Th:CaF\textsubscript{2} (figure~\ref{fig:229abs}) some common features can be detected. Similar concentration does not give similar absorption and the absorption profile is flat. The measurement of the crystal grown with the highest activity is uncertain due to its strong absorption. Following the spectral absorption of the other crystals we can expect its profile to be flat as well with a relative absorption higher than 6\,mm\textsuperscript{-1}. From figure~\ref{fig:229abs} it can be concluded that crystals grown using similar activity have similar absorption and at these low concentrations (compared to figure~\ref{fig:thabs}) the absorption does not heavily depend on concentration. \par

\begin{figure}[h!]
    \centering
    \includegraphics[width=\linewidth]{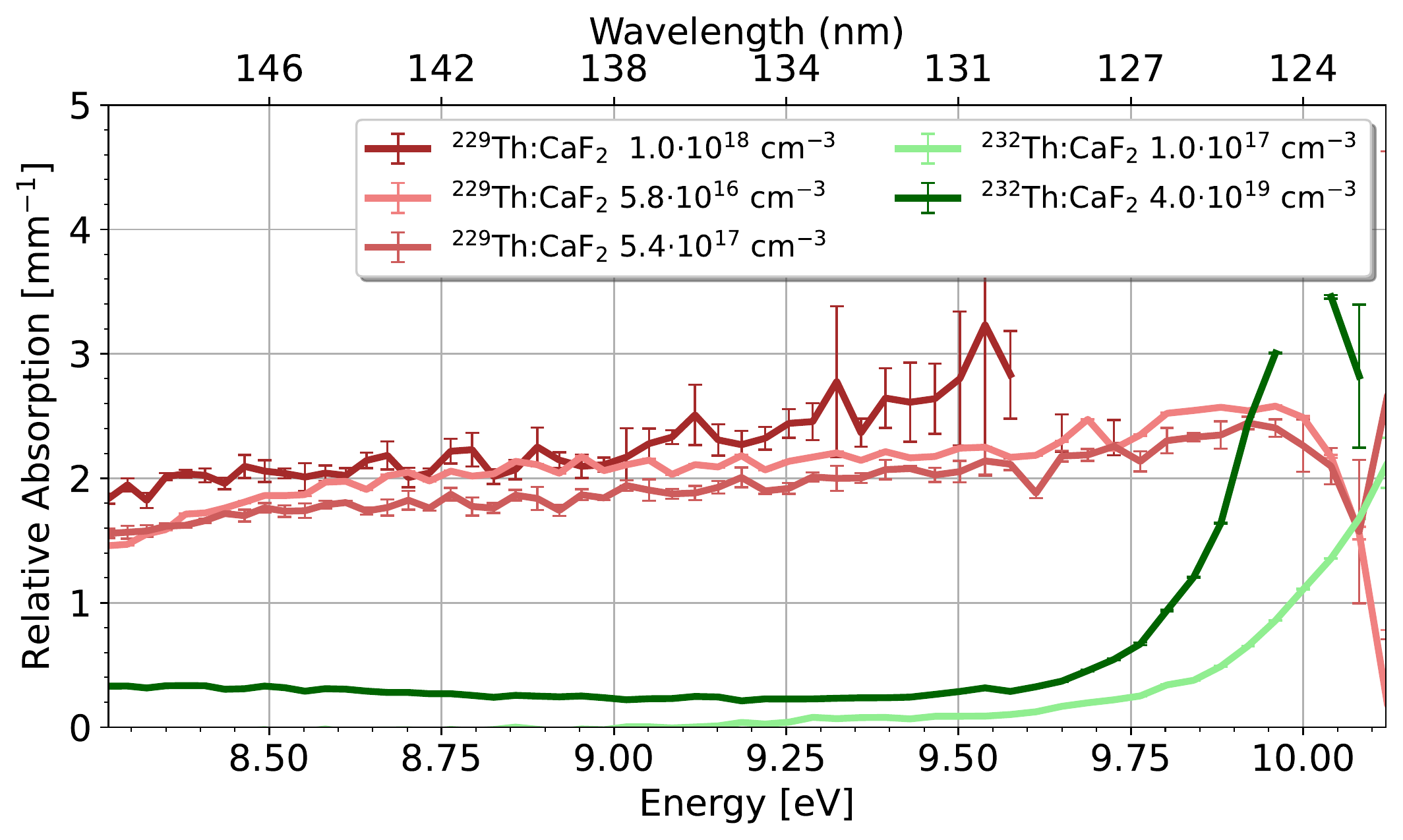}
    \caption{Relative absorption coefficient of Th:CaF\textsubscript{2} and undoped CaF\textsubscript{2} crystals with different concentrations for different energies and wavelengths. In the legend the crystal number, doping concentration and thickness is indicated. The highly doped $^{232}$Th doped crystal was made by manually mixing ThF\textsubscript{4} with CaF\textsubscript{2}, not by coprecipitation as described above. The differences in absorption between radioactive and not radioactive are clearly seen throughout the spectrum.}
    \label{fig:229232abs}
\end{figure}

The \textsuperscript{229}Th doped crystals display a different behavior from the \textsuperscript{232}Th doped crystals, which can be clearly seen when comparing figure~\ref{fig:thvuv} and \ref{fig:229vuv}. This was not expected as it is assumed that different isotopes behave identically concerning the electronic interactions. Most probably, the isotope itself does not change the characteristics but the radioactivity of the isotope does. The \textsuperscript{229}Th doped crystals only stop transmitting at the bandgap edge of CaF\textsubscript{2}, the 125\,nm absorption is not observed. The absorption around 150\,nm seems to have disappeared. The general transmission of the \textsuperscript{229}Th doped crystals is lower, despite the lower doping concentration. \par

To confirm the different absorption spectra of \textsuperscript{229}Th and \textsuperscript{232}Th doped crystals, drifts and changes between measurements of the VUV transmission setup were excluded. For this, a low doped \textsuperscript{232}Th:CaF\textsubscript{2}, a high doped \textsuperscript{232}Th:CaF\textsubscript{2} and a high doped \textsuperscript{229}Th:CaF\textsubscript{2} were measured in the same experimental run. The results are seen in figure~\ref{fig:229232abs}. Here we can see several effects. There seems to be an activity-related broadband VUV absorption, independent of doping concentration. The \textsuperscript{229}Th:CaF\textsubscript{2}, just as undoped CaF\textsubscript{2}, heavily absorbs at $>$10\,eV, the transmission edge of CaF\textsubscript{2}. The \textsuperscript{232}Th:CaF\textsubscript{2} absorbs heavily at $>$9.5\,eV, earlier than the other crystals. The intensity of this absorption is concentration-dependent, whereas the absorption of the radioactive crystals is almost concentration independent.

If we now combine the information contained in all figures, a few observations can be made: 

\begin{itemize}
    \item Crystals grown with increasing activity have an activity-dependent broadband VUV absorption, independent of Th concentration. This is clearly seen in the absorption of \textsuperscript{229}Th doped crystals and even in \textsuperscript{232}Th doped crystals probably due to the small amounts of added \textsuperscript{229}Th in combination with the weak activity of \textsuperscript{232}Th. The activity of the crystal with the highest \textsuperscript{232}Th doping concentration is still only 1\,Bq.
    \item Thorium doping with low activity creates a concentration-dependent absorption around 150\,nm and a strong absorption around 122\,nm. This is mainly visible for crystals with doping concentration of $>$7$\cdot$10\textsuperscript{19}\,cm\textsuperscript{-3}.
\end{itemize}

The thorium-related absorption around 122\,nm does not seem to be present in crystals grown with high activities. It is conjectured, that the thorium is in a different electronic state, either oxidation state or paired with a defect, if a crystal is grown in the presence of high radioactivity. Because of this change in electronic state, the absorption around 122\,nm disappears. The hypothesis connected with this observation is that the radioactivity induces loss of fluoride which produces non-stoichiometric, or fluoride-deficient crystals. \par

\begin{figure}[h!]
    \centering
    \includegraphics[width=0.5\linewidth]{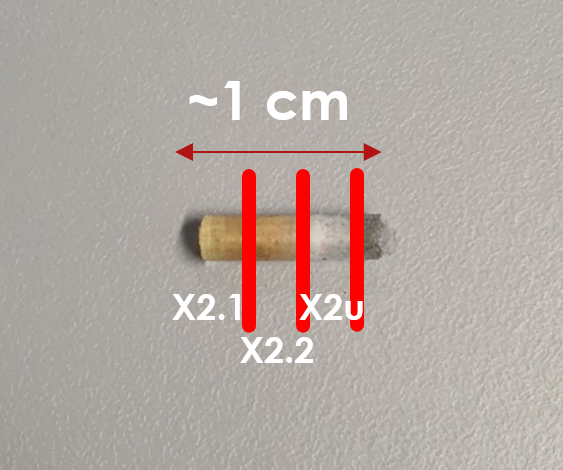}
    \caption{Cuts made in the crystal grown with 1\,MBq of $^{229}$Th, named X2. As can be seen from the orange coloring due to radioactively induced damage, only X2.1 and X2.2 are doped with $^{229}$Th.}
    \label{fig:x2cut}
\end{figure}

\subsection{Fluoride deficiency}
The strongest evidence for a fluoride deficiency in the \textsuperscript{229}Th:CaF\textsubscript{2} crystals was the transmission of the crystal named X2, containing a high activity of 1\,MBq. A thin slice taken from the undoped side of X2, named X2u (see figure~\ref{fig:x2cut}), was measured to be completely VUV opaque. At first the suspicion was radiation damage: Annealing to 600\,\textcelsius{} did not decrease VUV absorption of any part of X2 whereas it should remove radiation damage. The annealing did however remove the orange color of parts X2.1 and X2.2 in figure~\ref{fig:x2cut}, the healing of F centers. Not responding to annealing indicated that the VUV opaqueness of X2u was not radiation induced damage in the crystal. Another observation hinting towards fluoride-deficiency was that when growing CaF\textsubscript{2} crystals in vacuum (doped or not) outgassing of F\textsubscript{2} was measured by a mass spectrometer.\par

\begin{figure}[h!]
    \centering
    \includegraphics[width=\linewidth]{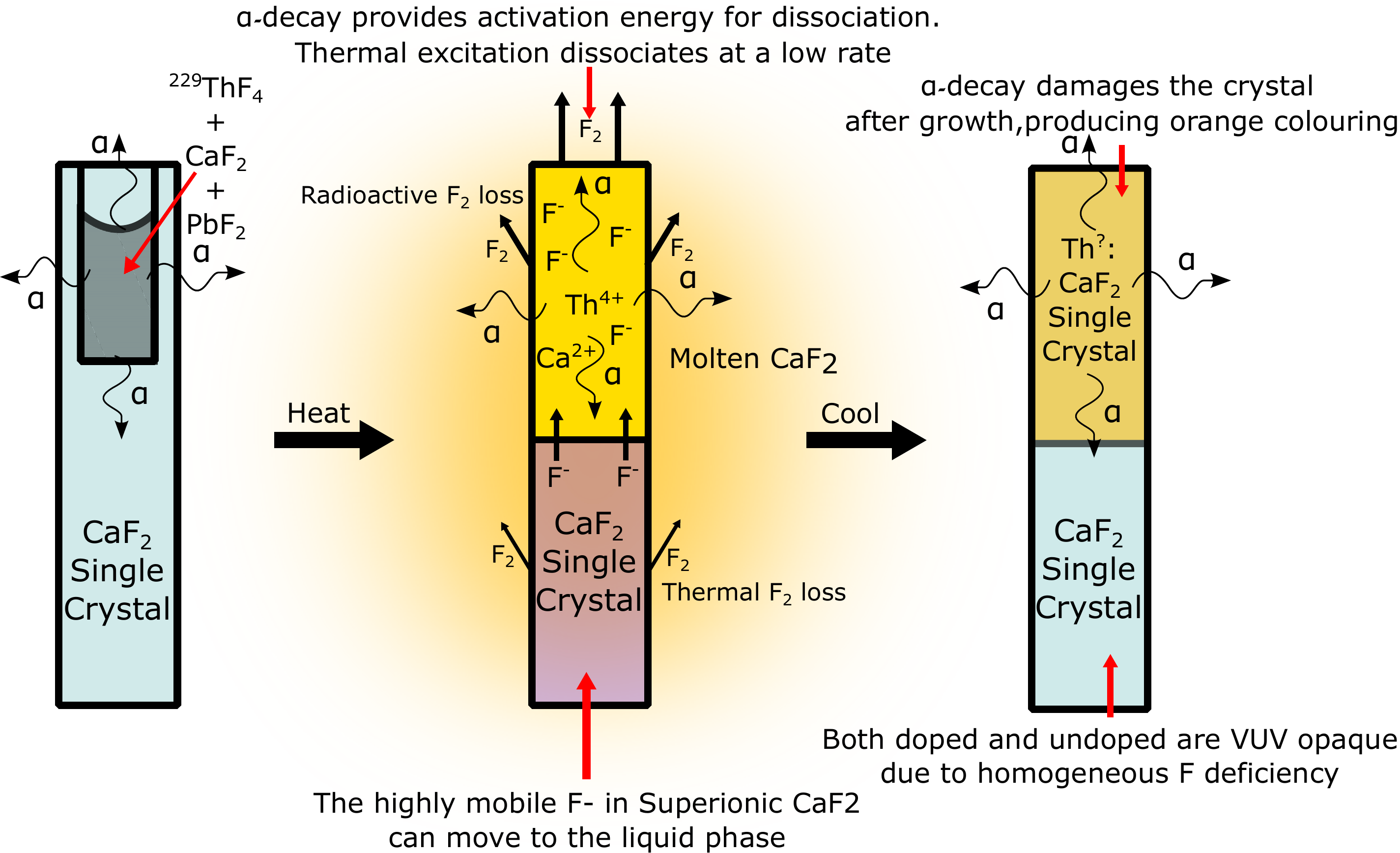}
    \caption{Schematic representation of the radioactively induced loss of fluoride during the growth process (compare to figure~\ref{fig:growthproc}), taking crystal X2 as an example (see figure~\ref{fig:x2cut}). First the crystal is filled with the radioactive powder. During growth, a part of the crystal is molten and a part of the crystal is in the superionic state. In the liquid, the \textsuperscript{229}Th is dissolved and the $\alpha$-decay locally provides the energy to dissociate CaF\textsubscript{2}, producing F\textsubscript{2} that is pumped away. The superionic crystal has extremely mobile F\textsuperscript{-} atoms which will diffuse to the liquid phase, supplying it with more F\textsuperscript{-}. The resulting crystal will have a homogeneous deficiency of F due to its high mobility during growth, but only a partial doping of Th due to its low mobility during growth. As is seen in figure~\ref{fig:c5}, only a small fraction of the Th penetrates the not molten crystal.}
    \label{fig:fdef}
\end{figure}

Our conjecture of the process that leads to the fluoride deficient crystals is depicted and detailed in figure~\ref{fig:fdef}. By growing CaF\textsubscript{2} with radioactive materials, the loss of fluoride in the liquid phase is enhanced and non-stoichiometric crystals are produced. The dissociation of fluoride compounds to produce gaseous F\textsubscript{2} through radioactivity, radiolysis, has been observed in UF\textsubscript{6} \cite{dmitrievskii1960dissociation}. The nuclear chemical reaction describing this process is  

\begin{equation}
    \text{CaF}_2 \xrightarrow[]{\alpha\text{ 5 MeV}} \text{Ca}^{s} + \text{F}_2^{g},
\label{eq:diss}
\end{equation}

where the metallic Ca is dissolved in the solid phase, and F\textsubscript{2} leaves the liquid melt as a gas. The energy for a chemical reaction is $\approx$\,eV, the energy of $\alpha$-decays is $\approx$\,MeV. Because of this, every single $\alpha$-decay could drive many chemical reactions. In a conservative estimate, if 1\,\% of all reactions with an $\alpha$ particle drive radiolysis, which costs 10\,eV per reaction, then 5000 dissociation events can take place per $\alpha$-decay. At an activity of 1 MBq of \textsuperscript{229}Th this would mean at least 5$\cdot10^9$ radiolysis events per second. In a growing cycle where a part of the crystal is molten for 22 hours, there are 2$\cdot10^{15}$ radiolysis events as compared to the $\approx10^{21}$ F atoms in the crystal. This would lead to a defect concentration of 4$\cdot10^{16}$\,cm\textsuperscript{-3}, which should indeed be visible in a transmission measurement. \par

Between 1370 K and 1450 K \cite{VORONIN20011349,DERRINGTON1975171}, the solid CaF$_2$ undergoes a phase transition into the superionic state. Therefore, the superionic part of the crystal becomes fluoride deficient due to fluoride transfer to the melt. In the superionic state, the fluorides are highly mobile inside the crystal Ca$^+$ matrix. The highly mobile F$^{-}$ can migrate from the superionic phase to the molten phase, where the radioactivity induces further dissociation. After freezing, the entire crystal is left fluoride deficient. The crystal compensates for F$^{-}$ loss by producing Ca metallic colloids which increase the VUV absorption. The top liquid thus loses fluoride supplied by the bottom solid. The mobile superionic fluorides evenly spread throughout the solid and liquid crystal, causing homogeneous fluoride-deficiency throughout the crystal explaining our observations. This fluoride deficiency causes a strong broadband absorption in the VUV spectral range. \par

\subsection{Cherenkov radiation}
A second characterization was performed where the radioluminescence of the $^{229}$Th:CaF\textsubscript{2} crystal, grown with 100 kBq of \textsuperscript{229}Th, was measured in the VUV spectral range. This crystal was chosen due to its high concentration and transparency. The inherent radioactivity of the $^{229}$Th produces two main types of luminescence: Luminescence of CaF$_2$ by creating electron-hole pairs which form self-trapped excitons (STE)~\cite{rodnyi1997physical} and Cherenkov radiation through beta emission with energies larger than 158\,keV. Stellmer et al.~\cite{Stellmer2016u233} measured both these emission types by using undoped CaF$_2$ and a solid $^{233}$U sample. The Cherenkov radiation dominates the low wavelength region up until 200\,nm. Above 200\,nm the luminescence of the STEs dominate. Both can produce a background for further experiments and thus merit characterization. 

\begin{figure}[h!]
\centering
\includegraphics[width=0.7\linewidth]{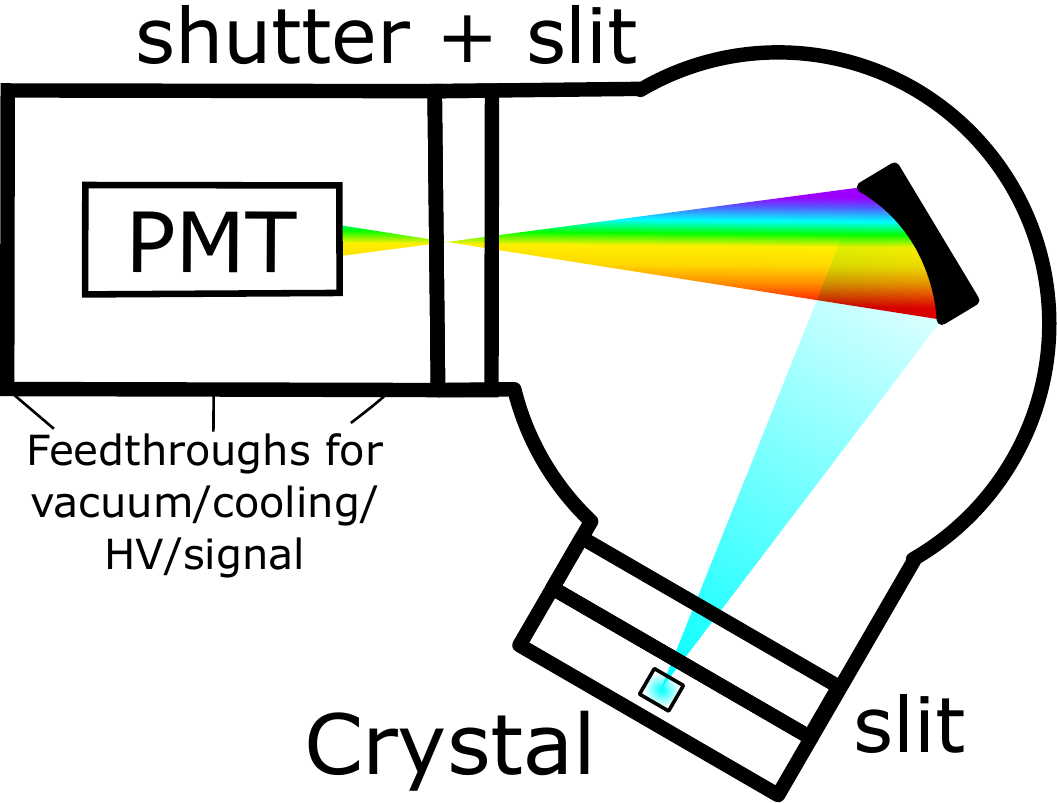}
\caption{Schematic drawing of the detector-spectrometer combination used to measure the fluorescence of activated crystals. The fluorescence produced by the crystal goes through the entrance slit and is then imaged by a focusing grating in a Seya-Namioka spectrometer model onto the exit slit. Only a slit size-dependent fraction of wavelengths will fall onto the PMT. Directly after the exit slit a fast (18.0\,ms opening time) shutter is installed.}
\label{spectro}
\end{figure}

\begin{figure}
\includegraphics[width=\linewidth]{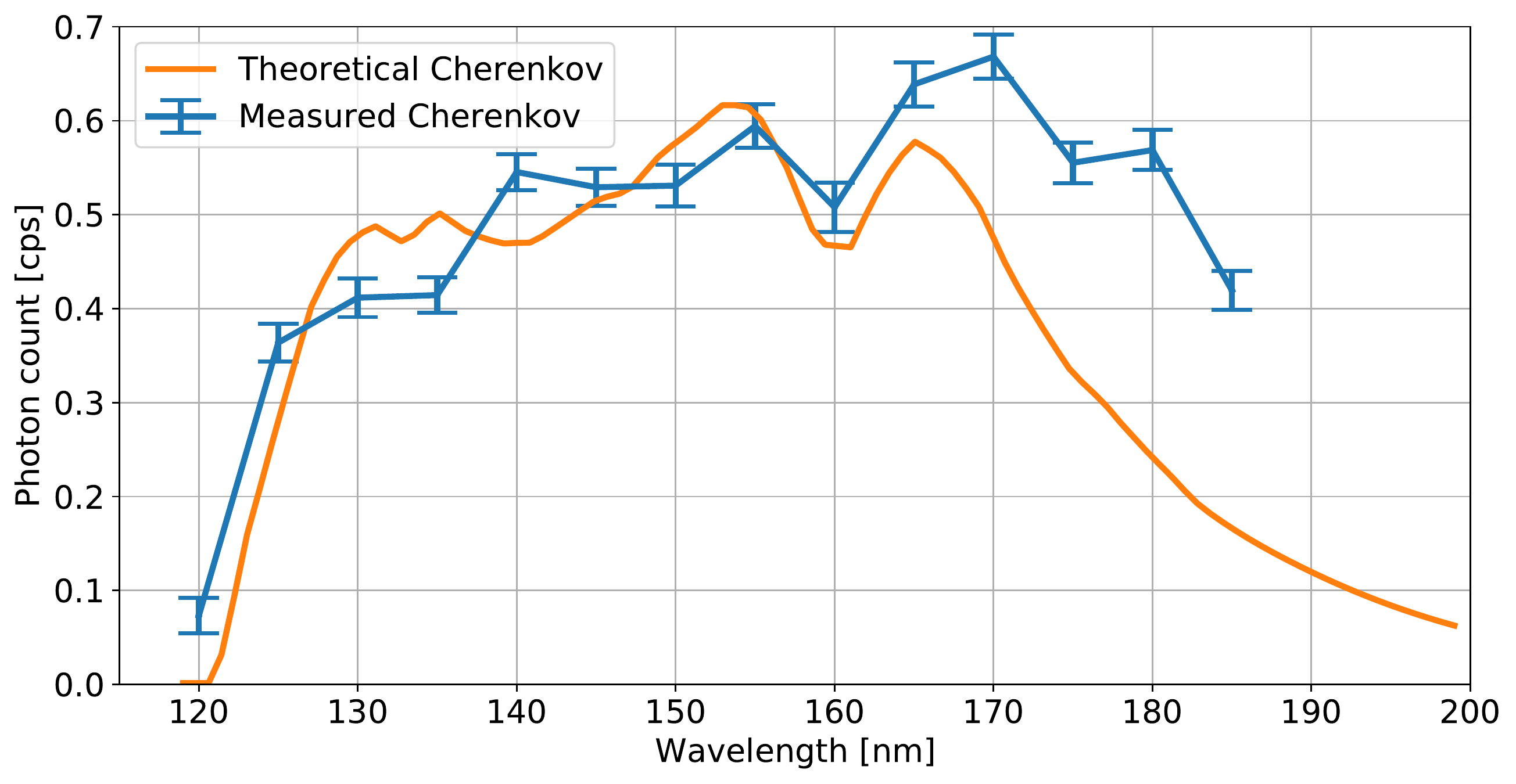}
\caption{The detected VUV spectrum of a $^{229}$Th:CaF$_2$ crystal compared to a theoretically computed Cherenkov spectrum that takes into account the spectral efficiency of the system (figure~\ref{spectro}). The calculation was slightly scaled to experimental results (see text for calculated discrepancy) to accentuate the differences between the curves. A 1\,mm slit size was used for the measurement. The maximum count rate was 0.7\,cps. The absorption edge of CaF$_2$ below 122\,nm can be seen, as well as loss of efficiency of the PMT and grating for higher wavelengths. The measured transmission of the crystal was used to calculate the theoretical Cherenkov spectrum. The sharp peaks at low wavelength in the theoretical spectrum are artefacts due to lines in the D2 lamp used to measure the transmission of this crystal. }
\label{Cherenkov}
\end{figure}

For the luminescence measurement, the chosen crystal was placed in the focal point of the entrance of a 234/302 McPherson spectrometer as depicted in figure~\ref{spectro}. Any radioluminescence was then spectrally resolved and imaged on a Hamamatsu R7639 PMT cooled to -30\,\textdegree C. The PMT is cooled to reduce dark noise to 0.5\,cps, otherwise Cherenkov radiation could not be detected.  A shutter was used to take continuous dark measurements while the signal was integrated for 9 hours per wavelength setting. The result of this measurement for a 1\,mm long 3.2\,mm diameter $^{229}$Th:CaF$_2$ crystal with a concentration of 5.4$\cdot10^{17}$\,cm$^{-3}$ can be seen in figure~\ref{Cherenkov}.

The Cherenkov background at 150\,nm was measured to be 0.5\,cps. As can be calculated~\cite{Stellmer2016u233}, every single decay event (1\,Bq) of $^{229}$Th produces 0.1 Cherenkov photon per second at 150\,nm in a 1\,nm bin. Considering 9.4\,kBq (the amount of nuclei visible through a narrow slit) of $^{229}$Th, and 0.05\,\% total efficiency of the spectrometer system, the Cherenkov flux at 150 nm would be 0.4\,cps. The small discrepancy between prediction and measurement can be due to the (lack of) accuracy in the measured amount of $^{229}$Th in the crystal, or the characterization of the numerous experimental efficiencies. In this spectral efficiency the measured transmission of the crystal, the efficiency of the grating with its MgF\textsubscript{2} coating and the efficiency of the PMT with its MgF\textsubscript{2} window were considered.  

 The calculated Cherenkov spectrum does not fully reproduce the measured values for higher wavelengths $>$170\,nm. This could be caused by measurement errors in the transmission of the crystal, which is used in the calculation of the spectrum. The mismatch can also be VUV luminescence of color centers induced by the radioactivity, radioluminescence. It is known that due to the scavenger, all crystals are slightly contaminated with Pb which has a luminescence peak at 180\,nm \cite{BABIN2004767}. 
 A more detailed investigation on the luminescence properties (radioluminescence and VUV-induced) is under way with the aim to identify contaminants as well as the additional electronic levels (defect centers) connected to Th doping in different charge states.

\subsection{Discussion and Conclusion}
We have grown highly doped ($5.4\cdot10^{17}$\,cm$^{-3}$) $^{229}$Th:CaF$_2$ crystals with low absorption ($\mu$ = 0.9\,mm$^{-1}$ or 20\,\% transmission at 150\,nm and 2\,mm thickness) and highly active (1\,MBq) $^{229}$Th:CaF$_2$. This was done by careful co-precipitation of ThF$_4$ and PbF$_2$ and mixing with CaF$_2$ to form a powder of growth material. The powder was then placed in a specially machined, millimeter-scale single CaF$_2$ seed crystal which was used to grow a single Th:CaF$_2$ crystal on top of the seed crystal using the vertical gradient freeze method. At high activities, these crystals seem to lose VUV transmission due to radioactivity-induced fluoride losses during the growth process (radiolysis). Even the not molten and undoped parts of this crystal lose transparency due to the high mobility of fluoride in the superionic state realized during crystal growth. Still, even at 1\,MBq total doping activity, a transmission of several percent remains. In the future, attempts will be made to regain the undoped CaF$_2$ transmission profile by re-feeding fluoride during growth or by annealing after growth.

 The thorium-related spectral absorption feature at 122\,nm decreased for radioactive crystals and the same crystals showed unidentified emission features. It is highly likely this is caused by a change in the electronic configuration of Th due to fluoride deficiency. By adding fluoride to the crystal it is expected that the electronic configuration changes to one similar to the non-radioactive crystals with high transparency. The highly active $^{229}$Th:CaF$_2$ crystals emit Cherenkov radiation which was quantified in this work, as it which will constitute a constant background for the future nuclear spectroscopy of $^{229}$Th in the crystal matrix. 

\subsection{Acknowledgements}
\begin{acknowledgments}
We acknowledge the tireless support for over a decade by Kurth Semmelroth, Rainer Apelt, and the scientific and technical staff of the Fraunhofer Institute for Integrated Systems and Device Technology IISB. We also thank Reinhard Uecker from the Institute for Crystal Growth for first explorations into Th:CaF$_2$ and invaluable advice. We are grateful to Mauro Tonelli from University of Pisa and the team from MEGA Materials for sharing their experience and insights.  

This work received funding from the European Research Council (ERC) under the European Union’s Horizon 2020 research and innovation programme under the ThoriumNuclearClock grant agreement No. 856415.  The research was supported by the Austrian Science Fund (FWF) Projects: I5971 (REThorIC) and P 33627 (NQRclock). 
\end{acknowledgments}
\bibliography{ms.bib}

\begin{thebibliography}{37}%
\makeatletter
\providecommand \@ifxundefined [1]{%
 \@ifx{#1\undefined}
}%
\providecommand \@ifnum [1]{%
 \ifnum #1\expandafter \@firstoftwo
 \else \expandafter \@secondoftwo
 \fi
}%
\providecommand \@ifx [1]{%
 \ifx #1\expandafter \@firstoftwo
 \else \expandafter \@secondoftwo
 \fi
}%
\providecommand \natexlab [1]{#1}%
\providecommand \enquote  [1]{``#1''}%
\providecommand \bibnamefont  [1]{#1}%
\providecommand \bibfnamefont [1]{#1}%
\providecommand \citenamefont [1]{#1}%
\providecommand \href@noop [0]{\@secondoftwo}%
\providecommand \href [0]{\begingroup \@sanitize@url \@href}%
\providecommand \@href[1]{\@@startlink{#1}\@@href}%
\providecommand \@@href[1]{\endgroup#1\@@endlink}%
\providecommand \@sanitize@url [0]{\catcode `\\12\catcode `\$12\catcode
  `\&12\catcode `\#12\catcode `\^12\catcode `\_12\catcode `\%12\relax}%
\providecommand \@@startlink[1]{}%
\providecommand \@@endlink[0]{}%
\providecommand \url  [0]{\begingroup\@sanitize@url \@url }%
\providecommand \@url [1]{\endgroup\@href {#1}{\urlprefix }}%
\providecommand \urlprefix  [0]{URL }%
\providecommand \Eprint [0]{\href }%
\providecommand \doibase [0]{http://dx.doi.org/}%
\providecommand \selectlanguage [0]{\@gobble}%
\providecommand \bibinfo  [0]{\@secondoftwo}%
\providecommand \bibfield  [0]{\@secondoftwo}%
\providecommand \translation [1]{[#1]}%
\providecommand \BibitemOpen [0]{}%
\providecommand \bibitemStop [0]{}%
\providecommand \bibitemNoStop [0]{.\EOS\space}%
\providecommand \EOS [0]{\spacefactor3000\relax}%
\providecommand \BibitemShut  [1]{\csname bibitem#1\endcsname}%
\let\auto@bib@innerbib\@empty
\bibitem [{\citenamefont {Von~der Wense}(2017)}]{von2017direct}%
  \BibitemOpen
  \bibfield  {author} {\bibinfo {author} {\bibfnamefont {L.}~\bibnamefont
  {Von~der Wense}},\ }\href@noop {} {\emph {\bibinfo {title} {On the direct
  detection of \textsuperscript{{229m}}Th}}}\ (\bibinfo  {publisher}
  {Springer},\ \bibinfo {year} {2017})\BibitemShut {NoStop}%
\bibitem [{\citenamefont {Tkalya}\ \emph {et~al.}(2015)\citenamefont {Tkalya},
  \citenamefont {Schneider}, \citenamefont {Jeet},\ and\ \citenamefont
  {Hudson}}]{Tkalya2015}%
  \BibitemOpen
  \bibfield  {author} {\bibinfo {author} {\bibfnamefont {E.~V.}\ \bibnamefont
  {Tkalya}}, \bibinfo {author} {\bibfnamefont {C.}~\bibnamefont {Schneider}},
  \bibinfo {author} {\bibfnamefont {J.}~\bibnamefont {Jeet}}, \ and\ \bibinfo
  {author} {\bibfnamefont {E.~R.}\ \bibnamefont {Hudson}},\ }\href {\doibase
  10.1103/PhysRevC.92.054324} {\bibfield  {journal} {\bibinfo  {journal} {Phys.
  Rev. C}\ }\textbf {\bibinfo {volume} {92}},\ \bibinfo {pages} {054324}
  (\bibinfo {year} {2015})}\BibitemShut {NoStop}%
\bibitem [{\citenamefont {Tkalya}(2011)}]{tkalya2011proposal}%
  \BibitemOpen
  \bibfield  {author} {\bibinfo {author} {\bibfnamefont {E.}~\bibnamefont
  {Tkalya}},\ }\href@noop {} {\bibfield  {journal} {\bibinfo  {journal}
  {Physical review letters}\ }\textbf {\bibinfo {volume} {106}},\ \bibinfo
  {pages} {162501} (\bibinfo {year} {2011})}\BibitemShut {NoStop}%
\bibitem [{\citenamefont {Borisyuk}\ \emph {et~al.}(2019)\citenamefont
  {Borisyuk}, \citenamefont {Kolachevsky}, \citenamefont {Taichenachev},
  \citenamefont {Tkalya}, \citenamefont {Tolstikhina},\ and\ \citenamefont
  {Yudin}}]{Borisyuk2019}%
  \BibitemOpen
  \bibfield  {author} {\bibinfo {author} {\bibfnamefont {P.~V.}\ \bibnamefont
  {Borisyuk}}, \bibinfo {author} {\bibfnamefont {N.~N.}\ \bibnamefont
  {Kolachevsky}}, \bibinfo {author} {\bibfnamefont {A.~V.}\ \bibnamefont
  {Taichenachev}}, \bibinfo {author} {\bibfnamefont {E.~V.}\ \bibnamefont
  {Tkalya}}, \bibinfo {author} {\bibfnamefont {I.~Y.}\ \bibnamefont
  {Tolstikhina}}, \ and\ \bibinfo {author} {\bibfnamefont {V.~I.}\ \bibnamefont
  {Yudin}},\ }\href {\doibase 10.1103/PhysRevC.100.044306} {\bibfield
  {journal} {\bibinfo  {journal} {Phys. Rev. C}\ }\textbf {\bibinfo {volume}
  {100}},\ \bibinfo {pages} {044306} (\bibinfo {year} {2019})}\BibitemShut
  {NoStop}%
\bibitem [{\citenamefont {Bilous}\ \emph {et~al.}(2017)\citenamefont {Bilous},
  \citenamefont {Kazakov}, \citenamefont {Moore}, \citenamefont {Schumm},\ and\
  \citenamefont {P\'alffy}}]{Bilous2017}%
  \BibitemOpen
  \bibfield  {author} {\bibinfo {author} {\bibfnamefont {P.~V.}\ \bibnamefont
  {Bilous}}, \bibinfo {author} {\bibfnamefont {G.~A.}\ \bibnamefont {Kazakov}},
  \bibinfo {author} {\bibfnamefont {I.~D.}\ \bibnamefont {Moore}}, \bibinfo
  {author} {\bibfnamefont {T.}~\bibnamefont {Schumm}}, \ and\ \bibinfo {author}
  {\bibfnamefont {A.}~\bibnamefont {P\'alffy}},\ }\href {\doibase
  10.1103/PhysRevA.95.032503} {\bibfield  {journal} {\bibinfo  {journal} {Phys.
  Rev. A}\ }\textbf {\bibinfo {volume} {95}},\ \bibinfo {pages} {032503}
  (\bibinfo {year} {2017})}\BibitemShut {NoStop}%
\bibitem [{\citenamefont {Nickerson}\ \emph {et~al.}(2020)\citenamefont
  {Nickerson}, \citenamefont {Pimon}, \citenamefont {Bilous}, \citenamefont
  {Gugler}, \citenamefont {Beeks}, \citenamefont {Sikorsky}, \citenamefont
  {Mohn}, \citenamefont {Schumm},\ and\ \citenamefont
  {P\'alffy}}]{Nickerson2020color}%
  \BibitemOpen
  \bibfield  {author} {\bibinfo {author} {\bibfnamefont {B.~S.}\ \bibnamefont
  {Nickerson}}, \bibinfo {author} {\bibfnamefont {M.}~\bibnamefont {Pimon}},
  \bibinfo {author} {\bibfnamefont {P.~V.}\ \bibnamefont {Bilous}}, \bibinfo
  {author} {\bibfnamefont {J.}~\bibnamefont {Gugler}}, \bibinfo {author}
  {\bibfnamefont {K.}~\bibnamefont {Beeks}}, \bibinfo {author} {\bibfnamefont
  {T.}~\bibnamefont {Sikorsky}}, \bibinfo {author} {\bibfnamefont
  {P.}~\bibnamefont {Mohn}}, \bibinfo {author} {\bibfnamefont {T.}~\bibnamefont
  {Schumm}}, \ and\ \bibinfo {author} {\bibfnamefont {A.}~\bibnamefont
  {P\'alffy}},\ }\href {\doibase 10.1103/PhysRevLett.125.032501} {\bibfield
  {journal} {\bibinfo  {journal} {Phys. Rev. Lett.}\ }\textbf {\bibinfo
  {volume} {125}},\ \bibinfo {pages} {032501} (\bibinfo {year}
  {2020})}\BibitemShut {NoStop}%
\bibitem [{\citenamefont {Nickerson}\ \emph {et~al.}(2021)\citenamefont
  {Nickerson}, \citenamefont {Pimon}, \citenamefont {Bilous}, \citenamefont
  {Gugler}, \citenamefont {Kazakov}, \citenamefont {Sikorsky}, \citenamefont
  {Beeks}, \citenamefont {Gr{\"u}neis}, \citenamefont {Schumm},\ and\
  \citenamefont {P{\'a}lffy}}]{nickerson2021driven}%
  \BibitemOpen
  \bibfield  {author} {\bibinfo {author} {\bibfnamefont {B.~S.}\ \bibnamefont
  {Nickerson}}, \bibinfo {author} {\bibfnamefont {M.}~\bibnamefont {Pimon}},
  \bibinfo {author} {\bibfnamefont {P.~V.}\ \bibnamefont {Bilous}}, \bibinfo
  {author} {\bibfnamefont {J.}~\bibnamefont {Gugler}}, \bibinfo {author}
  {\bibfnamefont {G.~A.}\ \bibnamefont {Kazakov}}, \bibinfo {author}
  {\bibfnamefont {T.}~\bibnamefont {Sikorsky}}, \bibinfo {author}
  {\bibfnamefont {K.}~\bibnamefont {Beeks}}, \bibinfo {author} {\bibfnamefont
  {A.}~\bibnamefont {Gr{\"u}neis}}, \bibinfo {author} {\bibfnamefont
  {T.}~\bibnamefont {Schumm}}, \ and\ \bibinfo {author} {\bibfnamefont
  {A.}~\bibnamefont {P{\'a}lffy}},\ }\href@noop {} {\bibfield  {journal}
  {\bibinfo  {journal} {Physical Review A}\ }\textbf {\bibinfo {volume}
  {103}},\ \bibinfo {pages} {053120} (\bibinfo {year} {2021})}\BibitemShut
  {NoStop}%
\bibitem [{\citenamefont {Tkalya}\ \emph {et~al.}(1996)\citenamefont {Tkalya},
  \citenamefont {Varlamov}, \citenamefont {Lomonosov},\ and\ \citenamefont
  {Nikulin}}]{tkalya1996processes}%
  \BibitemOpen
  \bibfield  {author} {\bibinfo {author} {\bibfnamefont {E.}~\bibnamefont
  {Tkalya}}, \bibinfo {author} {\bibfnamefont {V.}~\bibnamefont {Varlamov}},
  \bibinfo {author} {\bibfnamefont {V.}~\bibnamefont {Lomonosov}}, \ and\
  \bibinfo {author} {\bibfnamefont {S.}~\bibnamefont {Nikulin}},\ }\href@noop
  {} {\bibfield  {journal} {\bibinfo  {journal} {Physica Scripta}\ }\textbf
  {\bibinfo {volume} {53}},\ \bibinfo {pages} {296} (\bibinfo {year}
  {1996})}\BibitemShut {NoStop}%
\bibitem [{\citenamefont {Dessovic}\ \emph {et~al.}(2014)\citenamefont
  {Dessovic}, \citenamefont {Mohn}, \citenamefont {Jackson}, \citenamefont
  {Winkler}, \citenamefont {Schreitl}, \citenamefont {Kazakov},\ and\
  \citenamefont {Schumm}}]{dessovic2014229thorium}%
  \BibitemOpen
  \bibfield  {author} {\bibinfo {author} {\bibfnamefont {P.}~\bibnamefont
  {Dessovic}}, \bibinfo {author} {\bibfnamefont {P.}~\bibnamefont {Mohn}},
  \bibinfo {author} {\bibfnamefont {R.}~\bibnamefont {Jackson}}, \bibinfo
  {author} {\bibfnamefont {G.}~\bibnamefont {Winkler}}, \bibinfo {author}
  {\bibfnamefont {M.}~\bibnamefont {Schreitl}}, \bibinfo {author}
  {\bibfnamefont {G.}~\bibnamefont {Kazakov}}, \ and\ \bibinfo {author}
  {\bibfnamefont {T.}~\bibnamefont {Schumm}},\ }\href@noop {} {\bibfield
  {journal} {\bibinfo  {journal} {Journal of Physics: Condensed Matter}\
  }\textbf {\bibinfo {volume} {26}},\ \bibinfo {pages} {105402} (\bibinfo
  {year} {2014})}\BibitemShut {NoStop}%
\bibitem [{\citenamefont {Seiferle}\ \emph {et~al.}(2019)\citenamefont
  {Seiferle}, \citenamefont {von~der Wense}, \citenamefont {Bilous},
  \citenamefont {Amersdorffer}, \citenamefont {Lemell}, \citenamefont
  {Libisch}, \citenamefont {Stellmer}, \citenamefont {Schumm}, \citenamefont
  {D{\"u}llmann}, \citenamefont {P{\'a}lffy},\ and\ \citenamefont
  {Thirolf}}]{Seiferle2019}%
  \BibitemOpen
  \bibfield  {author} {\bibinfo {author} {\bibfnamefont {B.}~\bibnamefont
  {Seiferle}}, \bibinfo {author} {\bibfnamefont {L.}~\bibnamefont {von~der
  Wense}}, \bibinfo {author} {\bibfnamefont {P.~V.}\ \bibnamefont {Bilous}},
  \bibinfo {author} {\bibfnamefont {I.}~\bibnamefont {Amersdorffer}}, \bibinfo
  {author} {\bibfnamefont {C.}~\bibnamefont {Lemell}}, \bibinfo {author}
  {\bibfnamefont {F.}~\bibnamefont {Libisch}}, \bibinfo {author} {\bibfnamefont
  {S.}~\bibnamefont {Stellmer}}, \bibinfo {author} {\bibfnamefont
  {T.}~\bibnamefont {Schumm}}, \bibinfo {author} {\bibfnamefont {C.~E.}\
  \bibnamefont {D{\"u}llmann}}, \bibinfo {author} {\bibfnamefont
  {A.}~\bibnamefont {P{\'a}lffy}}, \ and\ \bibinfo {author} {\bibfnamefont
  {P.~G.}\ \bibnamefont {Thirolf}},\ }\href {\doibase
  10.1038/s41586-019-1533-4} {\bibfield  {journal} {\bibinfo  {journal}
  {Nature}\ }\textbf {\bibinfo {volume} {573}},\ \bibinfo {pages} {243}
  (\bibinfo {year} {2019})}\BibitemShut {NoStop}%
\bibitem [{\citenamefont {Sikorsky}\ \emph {et~al.}(2020)\citenamefont
  {Sikorsky}, \citenamefont {Geist}, \citenamefont {Hengstler}, \citenamefont
  {Kempf}, \citenamefont {Gastaldo}, \citenamefont {Enss}, \citenamefont
  {Mokry}, \citenamefont {Runke}, \citenamefont {D{\"u}llmann}, \citenamefont
  {Wobrauschek} \emph {et~al.}}]{sikorsky2020measurement}%
  \BibitemOpen
  \bibfield  {author} {\bibinfo {author} {\bibfnamefont {T.}~\bibnamefont
  {Sikorsky}}, \bibinfo {author} {\bibfnamefont {J.}~\bibnamefont {Geist}},
  \bibinfo {author} {\bibfnamefont {D.}~\bibnamefont {Hengstler}}, \bibinfo
  {author} {\bibfnamefont {S.}~\bibnamefont {Kempf}}, \bibinfo {author}
  {\bibfnamefont {L.}~\bibnamefont {Gastaldo}}, \bibinfo {author}
  {\bibfnamefont {C.}~\bibnamefont {Enss}}, \bibinfo {author} {\bibfnamefont
  {C.}~\bibnamefont {Mokry}}, \bibinfo {author} {\bibfnamefont
  {J.}~\bibnamefont {Runke}}, \bibinfo {author} {\bibfnamefont {C.~E.}\
  \bibnamefont {D{\"u}llmann}}, \bibinfo {author} {\bibfnamefont
  {P.}~\bibnamefont {Wobrauschek}},  \emph {et~al.},\ }\href@noop {} {\bibfield
   {journal} {\bibinfo  {journal} {Physical Review Letters}\ }\textbf {\bibinfo
  {volume} {125}},\ \bibinfo {pages} {142503} (\bibinfo {year}
  {2020})}\BibitemShut {NoStop}%
\bibitem [{\citenamefont {von~der Wense}\ \emph {et~al.}(2016)\citenamefont
  {von~der Wense}, \citenamefont {Seiferle}, \citenamefont {Laatiaoui},
  \citenamefont {Neumayr}, \citenamefont {Maier}, \citenamefont {Wirth},
  \citenamefont {Mokry}, \citenamefont {Runke}, \citenamefont {Eberhardt},
  \citenamefont {D{\"{u}}llmann}, \citenamefont {Trautmann},\ and\
  \citenamefont {Thirolf}}]{VonderWense2016}%
  \BibitemOpen
  \bibfield  {author} {\bibinfo {author} {\bibfnamefont {L.}~\bibnamefont
  {von~der Wense}}, \bibinfo {author} {\bibfnamefont {B.}~\bibnamefont
  {Seiferle}}, \bibinfo {author} {\bibfnamefont {M.}~\bibnamefont {Laatiaoui}},
  \bibinfo {author} {\bibfnamefont {J.~B.}\ \bibnamefont {Neumayr}}, \bibinfo
  {author} {\bibfnamefont {H.-J.}\ \bibnamefont {Maier}}, \bibinfo {author}
  {\bibfnamefont {H.-F.}\ \bibnamefont {Wirth}}, \bibinfo {author}
  {\bibfnamefont {C.}~\bibnamefont {Mokry}}, \bibinfo {author} {\bibfnamefont
  {J.}~\bibnamefont {Runke}}, \bibinfo {author} {\bibfnamefont
  {K.}~\bibnamefont {Eberhardt}}, \bibinfo {author} {\bibfnamefont {C.~E.}\
  \bibnamefont {D{\"{u}}llmann}}, \bibinfo {author} {\bibfnamefont {N.~G.}\
  \bibnamefont {Trautmann}}, \ and\ \bibinfo {author} {\bibfnamefont {P.~G.}\
  \bibnamefont {Thirolf}},\ }\href {\doibase 10.1038/nature17669} {\bibfield
  {journal} {\bibinfo  {journal} {Nature}\ }\textbf {\bibinfo {volume} {533}},\
  \bibinfo {pages} {47} (\bibinfo {year} {2016})}\BibitemShut {NoStop}%
\bibitem [{\citenamefont {Hehlen}\ \emph {et~al.}(2013)\citenamefont {Hehlen},
  \citenamefont {Greco}, \citenamefont {Rellergert}, \citenamefont {Sullivan},
  \citenamefont {DeMille}, \citenamefont {Jackson}, \citenamefont {Hudson},\
  and\ \citenamefont {Torgerson}}]{HEHLEN201391}%
  \BibitemOpen
  \bibfield  {author} {\bibinfo {author} {\bibfnamefont {M.~P.}\ \bibnamefont
  {Hehlen}}, \bibinfo {author} {\bibfnamefont {R.~R.}\ \bibnamefont {Greco}},
  \bibinfo {author} {\bibfnamefont {W.~G.}\ \bibnamefont {Rellergert}},
  \bibinfo {author} {\bibfnamefont {S.~T.}\ \bibnamefont {Sullivan}}, \bibinfo
  {author} {\bibfnamefont {D.}~\bibnamefont {DeMille}}, \bibinfo {author}
  {\bibfnamefont {R.~A.}\ \bibnamefont {Jackson}}, \bibinfo {author}
  {\bibfnamefont {E.~R.}\ \bibnamefont {Hudson}}, \ and\ \bibinfo {author}
  {\bibfnamefont {J.~R.}\ \bibnamefont {Torgerson}},\ }\href {\doibase
  https://doi.org/10.1016/j.jlumin.2011.09.037} {\bibfield  {journal} {\bibinfo
   {journal} {Journal of Luminescence}\ }\textbf {\bibinfo {volume} {133}},\
  \bibinfo {pages} {91} (\bibinfo {year} {2013})},\ \bibinfo {note} {16th
  International Conference on Luminescence ICL'11}\BibitemShut {NoStop}%
\bibitem [{\citenamefont {Rellergert}\ \emph {et~al.}(2010)\citenamefont
  {Rellergert}, \citenamefont {Sullivan}, \citenamefont {DeMille},
  \citenamefont {Greco}, \citenamefont {Hehlen}, \citenamefont {Jackson},
  \citenamefont {Torgerson},\ and\ \citenamefont {Hudson}}]{Rellergert_2010}%
  \BibitemOpen
  \bibfield  {author} {\bibinfo {author} {\bibfnamefont {W.~G.}\ \bibnamefont
  {Rellergert}}, \bibinfo {author} {\bibfnamefont {S.~T.}\ \bibnamefont
  {Sullivan}}, \bibinfo {author} {\bibfnamefont {D.}~\bibnamefont {DeMille}},
  \bibinfo {author} {\bibfnamefont {R.~R.}\ \bibnamefont {Greco}}, \bibinfo
  {author} {\bibfnamefont {M.~P.}\ \bibnamefont {Hehlen}}, \bibinfo {author}
  {\bibfnamefont {R.~A.}\ \bibnamefont {Jackson}}, \bibinfo {author}
  {\bibfnamefont {J.~R.}\ \bibnamefont {Torgerson}}, \ and\ \bibinfo {author}
  {\bibfnamefont {E.~R.}\ \bibnamefont {Hudson}},\ }\href {\doibase
  10.1088/1757-899x/15/1/012005} {\bibfield  {journal} {\bibinfo  {journal}
  {{IOP} Conference Series: Materials Science and Engineering}\ }\textbf
  {\bibinfo {volume} {15}},\ \bibinfo {pages} {012005} (\bibinfo {year}
  {2010})}\BibitemShut {NoStop}%
\bibitem [{\citenamefont {Rubloff}(1972)}]{caf2bandgap1976}%
  \BibitemOpen
  \bibfield  {author} {\bibinfo {author} {\bibfnamefont {G.~W.}\ \bibnamefont
  {Rubloff}},\ }\href {\doibase 10.1103/PhysRevB.5.662} {\bibfield  {journal}
  {\bibinfo  {journal} {Phys. Rev. B}\ }\textbf {\bibinfo {volume} {5}},\
  \bibinfo {pages} {662} (\bibinfo {year} {1972})}\BibitemShut {NoStop}%
\bibitem [{\citenamefont {Letz}\ \emph {et~al.}(2009)\citenamefont {Letz},
  \citenamefont {Gottwald}, \citenamefont {Richter},\ and\ \citenamefont
  {Parthier}}]{CaF2opticaltransmission}%
  \BibitemOpen
  \bibfield  {author} {\bibinfo {author} {\bibfnamefont {M.}~\bibnamefont
  {Letz}}, \bibinfo {author} {\bibfnamefont {A.}~\bibnamefont {Gottwald}},
  \bibinfo {author} {\bibfnamefont {M.}~\bibnamefont {Richter}}, \ and\
  \bibinfo {author} {\bibfnamefont {L.}~\bibnamefont {Parthier}},\ }\href
  {\doibase 10.1103/PhysRevB.79.195112} {\bibfield  {journal} {\bibinfo
  {journal} {Phys. Rev. B}\ }\textbf {\bibinfo {volume} {79}},\ \bibinfo
  {pages} {195112} (\bibinfo {year} {2009})}\BibitemShut {NoStop}%
\bibitem [{\citenamefont {Kazakov}\ \emph {et~al.}(2012)\citenamefont
  {Kazakov}, \citenamefont {Litvinov}, \citenamefont {Romanenko}, \citenamefont
  {Yatsenko}, \citenamefont {Romanenko}, \citenamefont {Schreitl},
  \citenamefont {Winkler},\ and\ \citenamefont {Schumm}}]{Kazakov_2012}%
  \BibitemOpen
  \bibfield  {author} {\bibinfo {author} {\bibfnamefont {G.~A.}\ \bibnamefont
  {Kazakov}}, \bibinfo {author} {\bibfnamefont {A.~N.}\ \bibnamefont
  {Litvinov}}, \bibinfo {author} {\bibfnamefont {V.~I.}\ \bibnamefont
  {Romanenko}}, \bibinfo {author} {\bibfnamefont {L.~P.}\ \bibnamefont
  {Yatsenko}}, \bibinfo {author} {\bibfnamefont {A.~V.}\ \bibnamefont
  {Romanenko}}, \bibinfo {author} {\bibfnamefont {M.}~\bibnamefont {Schreitl}},
  \bibinfo {author} {\bibfnamefont {G.}~\bibnamefont {Winkler}}, \ and\
  \bibinfo {author} {\bibfnamefont {T.}~\bibnamefont {Schumm}},\ }\href
  {\doibase 10.1088/1367-2630/14/8/083019} {\bibfield  {journal} {\bibinfo
  {journal} {New Journal of Physics}\ }\textbf {\bibinfo {volume} {14}},\
  \bibinfo {pages} {083019} (\bibinfo {year} {2012})}\BibitemShut {NoStop}%
\bibitem [{\citenamefont {Masuda}\ \emph {et~al.}(2019)\citenamefont {Masuda},
  \citenamefont {Yoshimi}, \citenamefont {Fujieda}, \citenamefont {Fujimoto},
  \citenamefont {Haba}, \citenamefont {Hara}, \citenamefont {Hiraki},
  \citenamefont {Kaino}, \citenamefont {Kasamatsu}, \citenamefont {Kitao},
  \citenamefont {Konashi}, \citenamefont {Miyamoto}, \citenamefont {Okai},
  \citenamefont {Okubo}, \citenamefont {Sasao}, \citenamefont {Seto},
  \citenamefont {Schumm}, \citenamefont {Shigekawa}, \citenamefont {Suzuki},
  \citenamefont {Stellmer}, \citenamefont {Tamasaku}, \citenamefont {Uetake},
  \citenamefont {Watanabe}, \citenamefont {Watanabe}, \citenamefont {Yasuda},
  \citenamefont {Yamaguchi}, \citenamefont {Yoda}, \citenamefont {Yokokita},
  \citenamefont {Yoshimura},\ and\ \citenamefont {Yoshimura}}]{Masuda2019}%
  \BibitemOpen
  \bibfield  {author} {\bibinfo {author} {\bibfnamefont {T.}~\bibnamefont
  {Masuda}}, \bibinfo {author} {\bibfnamefont {A.}~\bibnamefont {Yoshimi}},
  \bibinfo {author} {\bibfnamefont {A.}~\bibnamefont {Fujieda}}, \bibinfo
  {author} {\bibfnamefont {H.}~\bibnamefont {Fujimoto}}, \bibinfo {author}
  {\bibfnamefont {H.}~\bibnamefont {Haba}}, \bibinfo {author} {\bibfnamefont
  {H.}~\bibnamefont {Hara}}, \bibinfo {author} {\bibfnamefont {T.}~\bibnamefont
  {Hiraki}}, \bibinfo {author} {\bibfnamefont {H.}~\bibnamefont {Kaino}},
  \bibinfo {author} {\bibfnamefont {Y.}~\bibnamefont {Kasamatsu}}, \bibinfo
  {author} {\bibfnamefont {S.}~\bibnamefont {Kitao}}, \bibinfo {author}
  {\bibfnamefont {K.}~\bibnamefont {Konashi}}, \bibinfo {author} {\bibfnamefont
  {Y.}~\bibnamefont {Miyamoto}}, \bibinfo {author} {\bibfnamefont
  {K.}~\bibnamefont {Okai}}, \bibinfo {author} {\bibfnamefont {S.}~\bibnamefont
  {Okubo}}, \bibinfo {author} {\bibfnamefont {N.}~\bibnamefont {Sasao}},
  \bibinfo {author} {\bibfnamefont {M.}~\bibnamefont {Seto}}, \bibinfo {author}
  {\bibfnamefont {T.}~\bibnamefont {Schumm}}, \bibinfo {author} {\bibfnamefont
  {Y.}~\bibnamefont {Shigekawa}}, \bibinfo {author} {\bibfnamefont
  {K.}~\bibnamefont {Suzuki}}, \bibinfo {author} {\bibfnamefont
  {S.}~\bibnamefont {Stellmer}}, \bibinfo {author} {\bibfnamefont
  {K.}~\bibnamefont {Tamasaku}}, \bibinfo {author} {\bibfnamefont
  {S.}~\bibnamefont {Uetake}}, \bibinfo {author} {\bibfnamefont
  {M.}~\bibnamefont {Watanabe}}, \bibinfo {author} {\bibfnamefont
  {T.}~\bibnamefont {Watanabe}}, \bibinfo {author} {\bibfnamefont
  {Y.}~\bibnamefont {Yasuda}}, \bibinfo {author} {\bibfnamefont
  {A.}~\bibnamefont {Yamaguchi}}, \bibinfo {author} {\bibfnamefont
  {Y.}~\bibnamefont {Yoda}}, \bibinfo {author} {\bibfnamefont {T.}~\bibnamefont
  {Yokokita}}, \bibinfo {author} {\bibfnamefont {M.}~\bibnamefont {Yoshimura}},
  \ and\ \bibinfo {author} {\bibfnamefont {K.}~\bibnamefont {Yoshimura}},\
  }\href {\doibase 10.1038/s41586-019-1542-3} {\bibfield  {journal} {\bibinfo
  {journal} {Nature}\ }\textbf {\bibinfo {volume} {573}},\ \bibinfo {pages}
  {238} (\bibinfo {year} {2019})}\BibitemShut {NoStop}%
\bibitem [{\citenamefont {Jeet}\ \emph {et~al.}(2015)\citenamefont {Jeet},
  \citenamefont {Schneider}, \citenamefont {Sullivan}, \citenamefont
  {Rellergert}, \citenamefont {Mirzadeh}, \citenamefont {Cassanho},
  \citenamefont {Jenssen}, \citenamefont {Tkalya},\ and\ \citenamefont
  {Hudson}}]{Jeet2015}%
  \BibitemOpen
  \bibfield  {author} {\bibinfo {author} {\bibfnamefont {J.}~\bibnamefont
  {Jeet}}, \bibinfo {author} {\bibfnamefont {C.}~\bibnamefont {Schneider}},
  \bibinfo {author} {\bibfnamefont {S.~T.}\ \bibnamefont {Sullivan}}, \bibinfo
  {author} {\bibfnamefont {W.~G.}\ \bibnamefont {Rellergert}}, \bibinfo
  {author} {\bibfnamefont {S.}~\bibnamefont {Mirzadeh}}, \bibinfo {author}
  {\bibfnamefont {A.}~\bibnamefont {Cassanho}}, \bibinfo {author}
  {\bibfnamefont {H.~P.}\ \bibnamefont {Jenssen}}, \bibinfo {author}
  {\bibfnamefont {E.~V.}\ \bibnamefont {Tkalya}}, \ and\ \bibinfo {author}
  {\bibfnamefont {E.~R.}\ \bibnamefont {Hudson}},\ }\href {\doibase
  10.1103/PhysRevLett.114.253001} {\bibfield  {journal} {\bibinfo  {journal}
  {Phys. Rev. Lett.}\ }\textbf {\bibinfo {volume} {114}},\ \bibinfo {pages}
  {253001} (\bibinfo {year} {2015})}\BibitemShut {NoStop}%
\bibitem [{\citenamefont {Stellmer}\ \emph {et~al.}(2018)\citenamefont
  {Stellmer}, \citenamefont {Kazakov}, \citenamefont {Schreitl}, \citenamefont
  {Kaser}, \citenamefont {Kolbe},\ and\ \citenamefont {Schumm}}]{Stellmer2018}%
  \BibitemOpen
  \bibfield  {author} {\bibinfo {author} {\bibfnamefont {S.}~\bibnamefont
  {Stellmer}}, \bibinfo {author} {\bibfnamefont {G.}~\bibnamefont {Kazakov}},
  \bibinfo {author} {\bibfnamefont {M.}~\bibnamefont {Schreitl}}, \bibinfo
  {author} {\bibfnamefont {H.}~\bibnamefont {Kaser}}, \bibinfo {author}
  {\bibfnamefont {M.}~\bibnamefont {Kolbe}}, \ and\ \bibinfo {author}
  {\bibfnamefont {T.}~\bibnamefont {Schumm}},\ }\href {\doibase
  10.1103/PhysRevA.97.062506} {\bibfield  {journal} {\bibinfo  {journal} {Phys.
  Rev. A}\ }\textbf {\bibinfo {volume} {97}},\ \bibinfo {pages} {062506}
  (\bibinfo {year} {2018})}\BibitemShut {NoStop}%
\bibitem [{\citenamefont {Stöber}(1924)}]{Stober1925}%
  \BibitemOpen
  \bibfield  {author} {\bibinfo {author} {\bibfnamefont {F.}~\bibnamefont
  {Stöber}},\ }\href {\doibase doi:10.1524/zkri.1924.61.1.299} {\bibfield
  {journal} {\bibinfo  {journal} {Zeitschrift für Kristallographie -
  Crystalline Materials}\ }\textbf {\bibinfo {volume} {61}},\ \bibinfo {pages}
  {299} (\bibinfo {year} {1924})}\BibitemShut {NoStop}%
\bibitem [{\citenamefont {Schreitl}(2016)}]{schreitl2016growth}%
  \BibitemOpen
  \bibfield  {author} {\bibinfo {author} {\bibfnamefont {M.}~\bibnamefont
  {Schreitl}},\ }\emph {\bibinfo {title} {Growth and characterization of
  (doped) calcium fluoride crystals for the nuclear spectroscopy of Th-229}},\
  \href@noop {} {Ph.D. thesis},\ \bibinfo  {school} {Wien} (\bibinfo {year}
  {2016})\BibitemShut {NoStop}%
\bibitem [{\citenamefont {Molchanov}\ \emph {et~al.}(2005)\citenamefont
  {Molchanov}, \citenamefont {Friedrich}, \citenamefont {Wehrhan},\ and\
  \citenamefont {M{\"u}ller}}]{molchanov2005study}%
  \BibitemOpen
  \bibfield  {author} {\bibinfo {author} {\bibfnamefont {A.}~\bibnamefont
  {Molchanov}}, \bibinfo {author} {\bibfnamefont {J.}~\bibnamefont
  {Friedrich}}, \bibinfo {author} {\bibfnamefont {G.}~\bibnamefont {Wehrhan}},
  \ and\ \bibinfo {author} {\bibfnamefont {G.}~\bibnamefont {M{\"u}ller}},\
  }\href@noop {} {\bibfield  {journal} {\bibinfo  {journal} {Journal of crystal
  growth}\ }\textbf {\bibinfo {volume} {273}},\ \bibinfo {pages} {629}
  (\bibinfo {year} {2005})}\BibitemShut {NoStop}%
\bibitem [{\citenamefont {Ko}\ \emph {et~al.}(2001)\citenamefont {Ko},
  \citenamefont {Tozawa}, \citenamefont {Yoshikawa}, \citenamefont {Inaba},
  \citenamefont {Shishido}, \citenamefont {Oba}, \citenamefont {Oyama},
  \citenamefont {Kuwabara},\ and\ \citenamefont {Fukuda}}]{ko2001czochralski}%
  \BibitemOpen
  \bibfield  {author} {\bibinfo {author} {\bibfnamefont {J.}~\bibnamefont
  {Ko}}, \bibinfo {author} {\bibfnamefont {S.}~\bibnamefont {Tozawa}}, \bibinfo
  {author} {\bibfnamefont {A.}~\bibnamefont {Yoshikawa}}, \bibinfo {author}
  {\bibfnamefont {K.}~\bibnamefont {Inaba}}, \bibinfo {author} {\bibfnamefont
  {T.}~\bibnamefont {Shishido}}, \bibinfo {author} {\bibfnamefont
  {T.}~\bibnamefont {Oba}}, \bibinfo {author} {\bibfnamefont {Y.}~\bibnamefont
  {Oyama}}, \bibinfo {author} {\bibfnamefont {T.}~\bibnamefont {Kuwabara}}, \
  and\ \bibinfo {author} {\bibfnamefont {T.}~\bibnamefont {Fukuda}},\
  }\href@noop {} {\bibfield  {journal} {\bibinfo  {journal} {Journal of crystal
  growth}\ }\textbf {\bibinfo {volume} {222}},\ \bibinfo {pages} {243}
  (\bibinfo {year} {2001})}\BibitemShut {NoStop}%
\bibitem [{\citenamefont {Haynes}\ \emph {et~al.}(2016)\citenamefont {Haynes},
  \citenamefont {Lide},\ and\ \citenamefont {Bruno}}]{haynes2016crc}%
  \BibitemOpen
  \bibfield  {author} {\bibinfo {author} {\bibfnamefont {W.~M.}\ \bibnamefont
  {Haynes}}, \bibinfo {author} {\bibfnamefont {D.~R.}\ \bibnamefont {Lide}}, \
  and\ \bibinfo {author} {\bibfnamefont {T.~J.}\ \bibnamefont {Bruno}},\
  }\href@noop {} {\emph {\bibinfo {title} {CRC handbook of chemistry and
  physics}}}\ (\bibinfo  {publisher} {CRC press},\ \bibinfo {year}
  {2016})\BibitemShut {NoStop}%
\bibitem [{\citenamefont {Capelli}\ \emph {et~al.}(2015)\citenamefont
  {Capelli}, \citenamefont {Benes}, \citenamefont {Raison}, \citenamefont
  {Beilmann}, \citenamefont {Kunzel},\ and\ \citenamefont
  {Konings}}]{capelli2015thermodynamic}%
  \BibitemOpen
  \bibfield  {author} {\bibinfo {author} {\bibfnamefont {E.}~\bibnamefont
  {Capelli}}, \bibinfo {author} {\bibfnamefont {O.}~\bibnamefont {Benes}},
  \bibinfo {author} {\bibfnamefont {P.}~\bibnamefont {Raison}}, \bibinfo
  {author} {\bibfnamefont {M.}~\bibnamefont {Beilmann}}, \bibinfo {author}
  {\bibfnamefont {C.}~\bibnamefont {Kunzel}}, \ and\ \bibinfo {author}
  {\bibfnamefont {R.}~\bibnamefont {Konings}},\ }\href@noop {} {\bibfield
  {journal} {\bibinfo  {journal} {Journal of Chemical \& Engineering Data}\
  }\textbf {\bibinfo {volume} {60}},\ \bibinfo {pages} {3166} (\bibinfo {year}
  {2015})}\BibitemShut {NoStop}%
\bibitem [{\citenamefont {Beeks}\ and\ \citenamefont
  {Schumm}(2022)}]{beeks2022nuclear}%
  \BibitemOpen
  \bibfield  {author} {\bibinfo {author} {\bibfnamefont {K.}~\bibnamefont
  {Beeks}}\ and\ \bibinfo {author} {\bibfnamefont {T.}~\bibnamefont {Schumm}},\
  }\emph {\bibinfo {title} {The Nuclear Excitation of Thorium-229 in the
  CaF$_2$ Environment}},\ \href@noop {} {Ph.D. thesis},\ \bibinfo  {school} {TU
  Wien} (\bibinfo {year} {2022})\BibitemShut {NoStop}%
\bibitem [{\citenamefont {Reiterov}\ \emph {et~al.}(1980)\citenamefont
  {Reiterov}, \citenamefont {Trofimova},\ and\ \citenamefont
  {Shishatskaya}}]{reiterov1980influence}%
  \BibitemOpen
  \bibfield  {author} {\bibinfo {author} {\bibfnamefont {V.}~\bibnamefont
  {Reiterov}}, \bibinfo {author} {\bibfnamefont {L.}~\bibnamefont {Trofimova}},
  \ and\ \bibinfo {author} {\bibfnamefont {L.}~\bibnamefont {Shishatskaya}},\
  }\href@noop {} {\bibfield  {journal} {\bibinfo  {journal} {SOV. J. OPT.
  TECH.}\ }\textbf {\bibinfo {volume} {47}},\ \bibinfo {pages} {284} (\bibinfo
  {year} {1980})}\BibitemShut {NoStop}%
\bibitem [{\citenamefont {Denks}\ \emph {et~al.}(2000)\citenamefont {Denks},
  \citenamefont {Savikhina},\ and\ \citenamefont
  {Nagirnyi}}]{denks2000dependence}%
  \BibitemOpen
  \bibfield  {author} {\bibinfo {author} {\bibfnamefont {V.}~\bibnamefont
  {Denks}}, \bibinfo {author} {\bibfnamefont {T.}~\bibnamefont {Savikhina}}, \
  and\ \bibinfo {author} {\bibfnamefont {V.}~\bibnamefont {Nagirnyi}},\
  }\href@noop {} {\bibfield  {journal} {\bibinfo  {journal} {Applied surface
  science}\ }\textbf {\bibinfo {volume} {158}},\ \bibinfo {pages} {301}
  (\bibinfo {year} {2000})}\BibitemShut {NoStop}%
\bibitem [{\citenamefont {Rix}\ \emph {et~al.}(2011)\citenamefont {Rix},
  \citenamefont {Natura}, \citenamefont {Loske}, \citenamefont {Letz},
  \citenamefont {Felser},\ and\ \citenamefont {Reichling}}]{rix2011formation}%
  \BibitemOpen
  \bibfield  {author} {\bibinfo {author} {\bibfnamefont {S.}~\bibnamefont
  {Rix}}, \bibinfo {author} {\bibfnamefont {U.}~\bibnamefont {Natura}},
  \bibinfo {author} {\bibfnamefont {F.}~\bibnamefont {Loske}}, \bibinfo
  {author} {\bibfnamefont {M.}~\bibnamefont {Letz}}, \bibinfo {author}
  {\bibfnamefont {C.}~\bibnamefont {Felser}}, \ and\ \bibinfo {author}
  {\bibfnamefont {M.}~\bibnamefont {Reichling}},\ }\href@noop {} {\bibfield
  {journal} {\bibinfo  {journal} {Applied Physics Letters}\ }\textbf {\bibinfo
  {volume} {99}},\ \bibinfo {pages} {261909} (\bibinfo {year}
  {2011})}\BibitemShut {NoStop}%
\bibitem [{\citenamefont {Angervaks}\ \emph {et~al.}(2018)\citenamefont
  {Angervaks}, \citenamefont {Veniaminov}, \citenamefont {Stolyarchuk},
  \citenamefont {Vasilev}, \citenamefont {Kudryavtseva}, \citenamefont
  {Fedorov},\ and\ \citenamefont {Ryskin}}]{Angervaks:18}%
  \BibitemOpen
  \bibfield  {author} {\bibinfo {author} {\bibfnamefont {A.~E.}\ \bibnamefont
  {Angervaks}}, \bibinfo {author} {\bibfnamefont {A.~V.}\ \bibnamefont
  {Veniaminov}}, \bibinfo {author} {\bibfnamefont {M.~V.}\ \bibnamefont
  {Stolyarchuk}}, \bibinfo {author} {\bibfnamefont {V.~E.}\ \bibnamefont
  {Vasilev}}, \bibinfo {author} {\bibfnamefont {I.}~\bibnamefont
  {Kudryavtseva}}, \bibinfo {author} {\bibfnamefont {P.~P.}\ \bibnamefont
  {Fedorov}}, \ and\ \bibinfo {author} {\bibfnamefont {A.~I.}\ \bibnamefont
  {Ryskin}},\ }\href {\doibase 10.1364/JOSAB.35.001288} {\bibfield  {journal}
  {\bibinfo  {journal} {J. Opt. Soc. Am. B}\ }\textbf {\bibinfo {volume}
  {35}},\ \bibinfo {pages} {1288} (\bibinfo {year} {2018})}\BibitemShut
  {NoStop}%
\bibitem [{\citenamefont {Dmitrievskii}\ and\ \citenamefont
  {Migachev}(1960)}]{dmitrievskii1960dissociation}%
  \BibitemOpen
  \bibfield  {author} {\bibinfo {author} {\bibfnamefont {V.}~\bibnamefont
  {Dmitrievskii}}\ and\ \bibinfo {author} {\bibfnamefont {A.}~\bibnamefont
  {Migachev}},\ }\href@noop {} {\bibfield  {journal} {\bibinfo  {journal}
  {Journal of Nuclear Energy. Part A. Reactor Science}\ }\textbf {\bibinfo
  {volume} {12}},\ \bibinfo {pages} {185} (\bibinfo {year} {1960})}\BibitemShut
  {NoStop}%
\bibitem [{\citenamefont {Voronin}\ and\ \citenamefont
  {Volkov}(2001)}]{VORONIN20011349}%
  \BibitemOpen
  \bibfield  {author} {\bibinfo {author} {\bibfnamefont {B.}~\bibnamefont
  {Voronin}}\ and\ \bibinfo {author} {\bibfnamefont {S.}~\bibnamefont
  {Volkov}},\ }\href {\doibase https://doi.org/10.1016/S0022-3697(01)00036-1}
  {\bibfield  {journal} {\bibinfo  {journal} {Journal of Physics and Chemistry
  of Solids}\ }\textbf {\bibinfo {volume} {62}},\ \bibinfo {pages} {1349}
  (\bibinfo {year} {2001})}\BibitemShut {NoStop}%
\bibitem [{\citenamefont {Derrington}\ \emph {et~al.}(1975)\citenamefont
  {Derrington}, \citenamefont {Lindner},\ and\ \citenamefont
  {O'Keeffe}}]{DERRINGTON1975171}%
  \BibitemOpen
  \bibfield  {author} {\bibinfo {author} {\bibfnamefont {C.}~\bibnamefont
  {Derrington}}, \bibinfo {author} {\bibfnamefont {A.}~\bibnamefont {Lindner}},
  \ and\ \bibinfo {author} {\bibfnamefont {M.}~\bibnamefont {O'Keeffe}},\
  }\href {\doibase https://doi.org/10.1016/0022-4596(75)90241-8} {\bibfield
  {journal} {\bibinfo  {journal} {Journal of Solid State Chemistry}\ }\textbf
  {\bibinfo {volume} {15}},\ \bibinfo {pages} {171} (\bibinfo {year}
  {1975})}\BibitemShut {NoStop}%
\bibitem [{\citenamefont {Rodnyi}(1997)}]{rodnyi1997physical}%
  \BibitemOpen
  \bibfield  {author} {\bibinfo {author} {\bibfnamefont {P.~A.}\ \bibnamefont
  {Rodnyi}},\ }\href@noop {} {\emph {\bibinfo {title} {Physical processes in
  inorganic scintillators}}},\ Vol.~\bibinfo {volume} {14}\ (\bibinfo
  {publisher} {CRC press},\ \bibinfo {year} {1997})\BibitemShut {NoStop}%
\bibitem [{\citenamefont {Stellmer}\ \emph {et~al.}(2016)\citenamefont
  {Stellmer}, \citenamefont {Schreitl}, \citenamefont {Kazakov}, \citenamefont
  {Sterba},\ and\ \citenamefont {Schumm}}]{Stellmer2016u233}%
  \BibitemOpen
  \bibfield  {author} {\bibinfo {author} {\bibfnamefont {S.}~\bibnamefont
  {Stellmer}}, \bibinfo {author} {\bibfnamefont {M.}~\bibnamefont {Schreitl}},
  \bibinfo {author} {\bibfnamefont {G.~A.}\ \bibnamefont {Kazakov}}, \bibinfo
  {author} {\bibfnamefont {J.~H.}\ \bibnamefont {Sterba}}, \ and\ \bibinfo
  {author} {\bibfnamefont {T.}~\bibnamefont {Schumm}},\ }\href {\doibase
  10.1103/PhysRevC.94.014302} {\bibfield  {journal} {\bibinfo  {journal} {Phys.
  Rev. C}\ }\textbf {\bibinfo {volume} {94}},\ \bibinfo {pages} {014302}
  (\bibinfo {year} {2016})}\BibitemShut {NoStop}%
\bibitem [{\citenamefont {Babin}\ \emph {et~al.}(2004)\citenamefont {Babin},
  \citenamefont {{D. Oskam}}, \citenamefont {Vergeer},\ and\ \citenamefont
  {Meijerink}}]{BABIN2004767}%
  \BibitemOpen
  \bibfield  {author} {\bibinfo {author} {\bibfnamefont {V.}~\bibnamefont
  {Babin}}, \bibinfo {author} {\bibfnamefont {K.}~\bibnamefont {{D. Oskam}}},
  \bibinfo {author} {\bibfnamefont {P.}~\bibnamefont {Vergeer}}, \ and\
  \bibinfo {author} {\bibfnamefont {A.}~\bibnamefont {Meijerink}},\ }\href
  {\doibase https://doi.org/10.1016/j.radmeas.2003.12.015} {\bibfield
  {journal} {\bibinfo  {journal} {Radiation Measurements}\ }\textbf {\bibinfo
  {volume} {38}},\ \bibinfo {pages} {767} (\bibinfo {year} {2004})},\ \bibinfo
  {note} {proceedings of the 5th European Conference on Luminescent Detectors
  and Transformers of Ionizing Radiation (LUMDETR 2003)}\BibitemShut {NoStop}%
\end{thebibliography}%
\end{document}